\documentclass[a4paper,aps,pre]{revtex4}
\usepackage{graphicx}

\begin{document}

\title{Computer Simulation of Quantum Dynamics in a Classical Spin Environment
}

\author{Alessandro Sergi }
\email{sergi@ukzn.ac.za}

\affiliation{School of Chemistry and Physics \\
University of KwaZulu-Natal \\
Private Bag X01 Scottsville \\
3209 Pietermaritzburg \\
and \\
National Institute for Theoretical Physics (NITheP) \\
KwaZulu-Natal, South Africa 
}

\begin{abstract}
In this paper a formalism for studying the dynamics of quantum systems
coupled to classical spin environments is reviewed.
The theory is based on generalized antisymmetric brackets
and naturally predicts open-path off-diagonal
geometric phases in the evolution of the density matrix.
It is shown that such geometric phases must also be considered
in the quantum-classical Liouville equation for a classical bath
with canonical phase space coordinates; this occurs whenever the adiabatics basis
is complex (as in the case of a magnetic field coupled to the quantum subsystem).
When the quantum subsystem is weakly coupled to the
spin environment, non-adiabatic transitions can be neglected
and one can construct an effective non-Markovian computer simulation scheme
for open quantum system dynamics in classical spin environments.
In order to tackle this case,
integration algorithms based on the symmetric Trotter factorization
of the classical-like spin propagator are derived. Such algorithms are
applied to a model comprising a quantum two-level system
coupled to a single classical spin in an external magnetic field.
Starting from an excited state, the population difference and the
coherences of this two-state model are simulated in time
while the dynamics of the classical spin is monitored in detail.
It is the author's opinion that the numerical evidence provided in this paper
is a first step toward developing the simulation of quantum dynamics
in classical spin environments into an effective tool.
In turn, the ability to simulate such a dynamics can have a
positive impact on various fields, among which, for example, nano-science.
\end{abstract}

\maketitle


\section{Introduction}

The computer simulation of systems of interest to nano-science
requires to consider in detail the environment
surrounding, for example, the quantum reactive centers~\cite{kapral},
the Josephson junctions~\cite{josephson} or the quantum dots~\cite{dots1,dots2,dots3}.
Environments can be represented either by means of bosonic
degrees of freedom~\cite{leggett} or by spinors~\cite{prokofev}
(or also by a combination of the two types of coordinates).
In practice, following accurately the dynamics of both the relevant system
and the surrounding environment (bath) leads to a theoretical/computational approach
that is complementary to the study of master 
equations~\cite{petruccione,ottinger,ottinger-qm,ottinger-qm2,ottinger-qm3,ottinger-qm4,ottinger-qm5}.

A scheme to perform the computer simulation of quantum systems coupled
to classical spin baths was introduced in~\cite{sergi-spin}.
The approach of Ref.~\cite{sergi-spin} can be classified within the
quantum-classical 
approximations~\cite{anderson,prezhdo,balescu,balescu2,mcquarrie}
to quantum dynamics since
it is based on a quantum-classical 
Liouville equation~\cite{qcl1,qcl2,qcl3,qcl4,qcl5,qcl6,ilya}
for a classical spin bath.
It is interesting that the formalism in~\cite{sergi-spin}
does not require to approximate the memory function of the environment 
since the bath degrees of freedom are described explicitly,
in the spirit of molecular dynamics simulations~\cite{allentildesley,frenkelsmit}.
From this point of view, the approach of Ref.~\cite{sergi-spin}
provides a non-Markovian route to
the simulation of quantum effects in classical spin baths.
It is worthy of note that the formalism in~\cite{sergi-spin} was naturally devised 
exploiting the mathematical structure provided by
generalized antisymmetric brackets~\cite{b3,b4,b-silurante}.
Such brackets have also been used
to formulate the statistical mechanics of systems with 
thermodynamic~\cite{b1,b2,sergi-pvg}
and holonomic constraints~\cite{holo1,holo2} in classical mechanics.

In the regime of weak coupling,
the adiabatic basis is particularly suited for numerical studies.
When such a basis is complex
(\emph{e.g.}, in the presence of magnetic dipoles and fields),
it has been shown that the formalism of Ref.~\cite{sergi-spin}
naturally predicts an open~\cite{gonzalo,truhlar} 
geometric phase~\cite{berry,qphases,mead,kuratsuji} 
in the evolution of the off-diagonal matrix elements of operators.

In this paper, three topics will be dealt with.
The first is that, when the adiabatic basis is complex, also
the quantum-classical Liouville equation 
in a classical bath with canonical (position/momentum) coordinates
possesses an open geometric phase term for the off-diagonal matrix elements.
The second is a brief review of the formalism for the quantum dynamics
of systems in classical spin baths, as introduced in Ref.~\cite{sergi-spin}.
The third is the explicit formulation of integration algorithms,
based on the symmetric Trotter factorization~\cite{respa,respa2,ribes} of
the classical spin propagator.
It is worth noting that the classical-like spin dynamics could
also have been integrated by means of the elegant measure-preserving
algorithms invented by Gregory Ezra~\cite{ezra,sergi-ezra,sergi-ezra2}.
In order to illustrate the algorithms, a specific model comprising
a quantum  two-level system coupled to a classical spin 
is studied and numerical results are reported.
It is the author's opinion that the numerical algorithms presented here
are a first significant step toward developing
the theory introduced in Ref.~\cite{sergi-spin}
into an effective tool for studying quantum nano-systems.

This paper is organized as follow. In Sec.~\ref{sec:qc} the formulation
of quantum-classical dynamics of quantum systems in environments
represented by canonically conjugate variables is summarized.
It is shown that, when the basis is complex, 
one has to consider a geometric phase in the evolution of the
off-diagonal matrix elements also in this case.
In Sec.~\ref{sec:qcsd} it is briefly reviewed how the quantum-classical
theory of Sec.~\ref{sec:qc} can be generalized to describe
quantum systems in classical spin baths.
The model system simulated in this paper is introduced in Sec.~\ref{sec:model}.
Its time-reversible integrators, on the various energy surfaces, are
developed in Sec.~\ref{sec:integrator}.
The detailed algorithms for integration on the $(1,1)$, $(2,2)$ and
$(1,2)$ surfaces are given in the Appendices~\ref{app:int-11}-\ref{app:int-12}.
The details and the results of the calculations on the model
are discussed in Sec.~\ref{sec:results}.
Finally, conclusions and perspectives are given in Sec.~\ref{sec:conc}.

\section{Quantum-classical Liouville equation in a complex adiabatic basis}
\label{sec:qc}
 
Consider a quantum system whose Hamiltonian operator $\hat{H}(\{\xi\})$
is defined in terms of a set of quantum operators $\hat{\xi}_i$, $i=1,\ldots,n$
and assume that the quantum degrees of freedom interact with a classical bath
which can be represented by canonical phase space coordinates $X=(R,P)$.
Such classical coordinates enter the definition of the bath classical Hamiltonian $H_B(X)$.
It is worth stating clearly that in this section a multidimensional notation is adopted.
According to this, for example, the symbol $R$ stands for $(R_1,R_2, ...)$, and
a scalar products such as $P\cdot P$ stand for $\sum_I P_I^2$.
Such a multidimensional notation will be abandoned in favor of a more
explicit notation in the next sections.
The interaction between the quantum subsystem and the classical bath
is given in terms of a coupling term $\hat{H}_C(\{\xi\},X)$.
The total Hamiltonian (which must be a constant of motion)
describing the coupled quantum subsystem
plus classical bath can be written as
\begin{equation}
\hat{\cal H}(X)=\hat{H}(\{\xi\})+\hat{H}_{\rm C}(\{\xi\},X)+H_{\rm B}(X)\;.
\label{eq:tot-ham}
\end{equation}
Accordingly, one is also led to introduce a quantum-classical
density matrix $\hat{\rho}_{\rm QC}(X)$ and quantum-classical
operators $\hat{\chi}_{\rm QC}(X)$.
The evolution in time, in a Schr\"odinger-like dynamical picture
can be postulated as
\begin{eqnarray}
\frac{\partial}{\partial t}\hat{\rho}_{\rm QC}(X,t)
&=&-\frac{i}{\hbar}\left[\begin{array}{cc} \hat{\cal H}(X) & 
\hat{\rho}_{\rm QC}(X,t)\end{array}\right]
\mbox{\boldmath$\cal D$}
\left[\begin{array}{c}\hat{\cal H}(X)\\ 
\hat{\rho}_{\rm QC}(X,t)\end{array}\right]
\;,\nonumber\\
&=&-\frac{i}{\hbar}\left[\hat{\cal H}(X),\hat{\rho}_{\rm QC}(X,t)
\right]_{\mbox{\tiny \boldmath$\cal D$}} \;.
\label{eq:pW}
\end{eqnarray}
In Equation~(\ref{eq:pW}) the antisymmetric matrix operator $\mbox{\boldmath$\cal D$}$
has been introduced. It is defined as
\begin{eqnarray}
\mbox{\boldmath$\cal D$}
&=&
\left[\begin{array}{cc} 0 & 1+\frac{i\hbar}{2}
\frac{\overleftarrow{\partial}}{\partial X_I}{\cal B}_{IJ}
\frac{\overrightarrow{\partial}}{\partial X_J}
\\
-1-\frac{i\hbar}{2}
\frac{\overleftarrow{\partial}}{\partial X_I}{\cal B}_{IJ}
\frac{\overrightarrow{\partial}}{\partial X_J}
& 0\end{array}\right]\;,
\nonumber\\
\end{eqnarray}
where
\begin{equation}
\mbox{\boldmath$\cal B$}=\left[\begin{array}{cc}0 & 1\\-1 & 0\end{array}\right]
\end{equation}
is the symplectic matrix.
The arrows over the partial derivative symbols in Eq.~(\ref{eq:pW}) denote
in which direction the partial derivative operator must act.
Moreover, in Eq.~(\ref{eq:pW}) and in the following the sum over repeated
indices is implied.
Equation~(\ref{eq:pW}), which is just the quantum-classical Liouville 
equation~\cite{qcl1,qcl2,qcl3,qcl4,qcl5,qcl6,ilya} written in matrix form, 
defines what is known as 
quantum-classical bracket~\cite{anderson,prezhdo,balescu,balescu2,mcquarrie}
or non-Hamiltonian commutator~\cite{b3}.
The bracket couples the dynamics of the phase space degrees of freedom
with that of the quantum operators; it takes into account
both the conservation of the energy and the quantum back-reaction.
Moreover, when there is no coupling, i.e., $\hat{H}_{\rm C}=0$,
the bracket makes the quantum system evolve in terms
of the standard quantum commutator and the classical bath through
the Poisson bracket.

The quantum-classical bracket in Eq.~(\ref{eq:pW})
does not satisfy the Jacobi relation and this, in turn,
leads to the lack of time-translation invariance
of the algebra defined in terms of the bracket itself~\cite{b3,nielsen}.
Less abstract consequences of such mathematical features are
that the coordinates of the bath, which are classical at time zero,
acquire quantum phases as time flows.
With respect to this,
one has to consider the bracket as an approximation
to the correct quantum dynamics of the total system (subsystem plus bath).
Such a dynamics, although correct in principle, would not be calculable
so that, following the philosophy of \emph{approximated theoretical complexity}
(as discussed in~\cite{pe}),
the quantum-classical bracket can be invoked as an effective
tool in order to perform computer simulations that would be otherwise impossible. 
In practice, fast bath decoherence may alleviate the theoretical problems
associated with the acquisition of quantum phase terms by the variables
that should stay classical and, indeed,
the quantum-classical bracket, or quantum-classical Liouville equation,
is used for many applications in chemistry and 
physics~\cite{anderson,prezhdo,balescu,balescu2,mcquarrie,qcl1,qcl2,qcl3,qcl4,qcl5,qcl6,ilya}.
It is worth reminding
that the non-Lie (or, as they are also called, non-Hamiltonian) brackets,
with their lack of time translation invariance,
are also used as technical tools to impose thermodynamical
(such as constant temperature and/or pressure)~\cite{b1,b2,sergi-pvg}
and holonomic constraints~\cite{holo1,holo2}
in classical molecular dynamics calculations~\cite{allentildesley,frenkelsmit}.

The abstract equations of motion in~({\ref{eq:pW}) can
be represented in the adiabatic basis. 
Upon writing the total Hamiltonian as $\hat{\cal H}(X)=(P^2/2M)+\hat{h}(R)$,
such a basis is defined by the eigenvalue problem
$\hat{h}(R)|\alpha;R\rangle=E_{\alpha}(R)|\alpha;R\rangle$.
In the adiabatic basis the quantum-classical evolution reads
\begin{eqnarray}
\frac{\partial}{\partial t}\hat{\rho}_{\alpha\alpha'}(X,t)
&=&-\sum_{\beta\beta'}\left[
i\omega_{\alpha\alpha'}+iL_{\alpha\alpha'}
\delta_{\alpha\beta}\delta_{\alpha'\beta'}
\right. \nonumber\\
&+&\left.J_{\alpha\alpha',\beta\beta'}\right]\hat{\rho}_{\beta\beta'}(X,t) \;.
\label{eq:qcle-ad}
\end{eqnarray}
In Equation~(\ref{eq:qcle-ad}) the symbol
$ \omega_{\alpha\alpha'}(R) =[E_{\alpha}(R)-E_{\alpha'}(R)]/\hbar$
denotes the Bohr frequency,
\begin{equation}
iL_{\alpha\alpha'}=\frac{P}{M}\cdot\frac{\partial}{\partial R}
+\frac{1}{2}\left(F_W^{\alpha}+F_W^{\alpha'}\right)\cdot
\frac{\partial}{\partial P}\;,
\end{equation}
is the classical-like Liouville operator for the bath degrees of freedom,
$F_W^{\alpha}=-\langle\alpha;R|(\partial\hat{h}/\partial R)|\alpha;R\rangle$
is the Helmann-Feynman force, 
and 
\begin{eqnarray}
J_{\alpha\alpha',\beta\beta'}
&=&
\frac{P}{M}\cdot d_{\alpha\beta}
\left(1+\frac{1}{2}S_{\alpha\beta}
\cdot\frac{\partial}{\partial P}\right)\delta_{\alpha'\beta'}
\nonumber\\
&+&
\frac{P}{M}\cdot d_{\alpha'\beta'}^*
\left(1+\frac{1}{2}S_{\alpha'\beta'}^*
\cdot\frac{\partial}{\partial P}\right)\delta_{\alpha\beta}\;,
\label{eq:TO}
\end{eqnarray}
is the transition operator,
responsible for the non-adiabatic transitions between the energy
levels of the quantum subsystem, as a result of the coupling to the bath.
The symbol $d_{\alpha\beta} = \langle\alpha|\partial/\partial R|
\beta\rangle$ denotes the non-adiabatic coupling vector.
In Equation~(\ref{eq:TO}),
the vector $S_{\alpha\beta}=\hat{d}_{\alpha\beta}
(E_{\alpha}-E_{\beta})
(\hat{d}_{\alpha\beta}\cdot P/M)^{-1}$ together with its complex conjugate
$S_{\alpha'\beta'}^*$ have been defined (the symbol $\hat{d}_{\alpha\beta}$
denotes the normalization of the coupling vector over the space
of all $R$ coordinates).

The coupling vector has the property $d_{\alpha\beta}=-d_{\beta\alpha}^*$
so that, when the adiabatic basis is real, $d_{\alpha\alpha}=0$
and the transition operator in eq.~(\ref{eq:TO}) is purely off-diagonal.
However, when the basis is complex (e.g., when the symmetry under time-inversion is broken,
for example, because of a magnetic field),
$d_{\alpha\alpha}$ is not zero and is purely imaginary
\begin{equation}
d_{\alpha\alpha}=i\phi_{\alpha\alpha}\ne 0\;.
\end{equation}
One can see that a phase $\phi_{\alpha\alpha}$ is naturally emerging
from the representation of the quantum-classical Liouville
equation in a complex basis.
Hence, in a complex adiabatic basis, Eq.~(\ref{eq:qcle-ad}) can be rewritten as
\begin{eqnarray}
\frac{\partial}{\partial t}\hat{\rho}_{\alpha\alpha'}(X,t)
&=&-\sum_{\beta\beta'}\Bigg[
i\omega_{\alpha\alpha'}
+i\frac{P}{M}\left(\phi_{\alpha\alpha}-\phi_{\alpha'\alpha'}\right)
\nonumber\\
&+&iL_{\alpha\alpha'}
\delta_{\alpha\beta}\delta_{\alpha'\beta'}
+J^{\rm od}_{\alpha\alpha',\beta\beta'}\Bigg]
\hat{\rho}_{\beta\beta'}(X,t) \;.
\nonumber\\
\label{eq:qcle-ad-complex}
\end{eqnarray}
where $J^{\rm od}_{\alpha\alpha',\beta\beta'}$ is the
off-diagonal part of the transition operator
\begin{eqnarray}
J^{\rm od}_{\alpha\alpha',\beta\beta'}
&=&
\frac{P}{M}\cdot d_{\alpha\beta}[1-\delta_{\alpha\beta}]
\left(1+\frac{1}{2}S_{\alpha\beta}
\cdot\frac{\partial}{\partial P}\right)\delta_{\alpha'\beta'}
\nonumber\\
&+&
\frac{P}{M}\cdot d_{\alpha'\beta'}^*[1-\delta_{\alpha'\beta'}]
\left(1+\frac{1}{2}S_{\alpha'\beta'}^*
\cdot\frac{\partial}{\partial P}\right)\delta_{\alpha\beta}\;,
\label{eq:TO-off}
\nonumber
\\
\end{eqnarray}
While the phase $\omega_{\alpha\alpha}$ has a dynamical source,
the phase $\phi_{\alpha\alpha}$ has a geometric origin
and it is analogous to the famous Berry phase~\cite{berry,qphases,mead}.
Hence, Equation~(\ref{eq:qcle-ad-complex}) is the quantum-classical Liouville
equation displaying geometric phase effects.
In principle, such phases are also present for open paths~\cite{pati}
of the classical environment and they are
off-diagonal in nature~\cite{filipp,englman,manini}.


\section{Quantum-Classical Spin Dynamics}
\label{sec:qcsd}

Consider a classical spin vector $\bf S$ with components 
$S_I$, $I=x,y,z$,
whose energy is described by the Hamiltonian be $H_{\rm SB}({\bf S})$.
It is known that  the equations of motion can be written in matrix form as
\begin{equation}
\dot{S}_I={\cal B}_{IJ}^{\rm S}\frac{\partial H_{\rm SB}}{\partial S_J}\;,
\label{eq:eqofm}
\end{equation}
where
\begin{equation}
\mbox{\boldmath $\cal B$}^{\rm S}
=\left[\begin{array}{ccc} 0   &  S_z & -S_y \\ 
                         -S_z &  0   &  S_x \\
                          S_y & -S_x &  0 \end{array}\right] 
\;.
\label{eq:bmat}
\end{equation}
The antisymmetric matrix $\mbox{\boldmath $\cal B$}^{\rm S}$ can also be written
in a compact way as
\begin{equation}
{\cal B}_{IJ}^{\rm S}=\epsilon_{IJK}S_K \;.
\end{equation}
The equations of motion~(\ref{eq:eqofm})
preserve the Casimir $C_2={\bf S}\cdot {\bf S}$
for any arbitrary Hamiltonian $H_{\rm SB}({\bf S})$.
They also have a zero phase space compressibility
\begin{eqnarray}
\kappa&=&\frac{\partial \dot{S}_I}{\partial S_I}
\epsilon_{IJK}\delta_{KJ}\frac{\partial H_{\rm SB}}{\partial S_J}
+{\cal B}_{IJ}^{\rm S}\frac{\partial^2 H_{\rm SB}}{\partial S_I\partial S_J}=0\;.
\end{eqnarray}
The equations of motion can also be written in the form
\begin{equation}
\dot{S}_I=\{S_I,H_{\rm SB}\}_{\tiny \mbox{\boldmath$\cal B$}^{\rm S}}\;,
\end{equation} 
upon introducing a non-canonical
bracket defined as
\begin{equation}
\{A,B\}_{\mbox{ \tiny \boldmath$\cal B$}^{\rm S}}
=\frac{\partial A}{\partial S_I}{\cal B}_{IJ}^{\rm S}
\frac{\partial B}{\partial S_J}\;,
\end{equation}
where $A=A({\bf S})$ and $B=B({\bf S})$ are arbitrary functions
of the spin degrees of freedom.

Let us assume that the classical spin system is interacting with 
the quantum system with Hamiltonian operator $\hat{H}(\{\hat{\chi}\})$
through an interaction of the form $\hat{H}_{\rm c}(\{\hat{\chi}\},{\bf S})$
The total Hamiltonian operator of the quantum subsystem
in the classical spin bath can be written analogously to
Eq.~(\ref{eq:tot-ham})
\begin{equation}
\hat{\cal H}({\bf S})=\hat{H}(\{\hat{\chi}\})
+\hat{H}_{\rm C}(\{\hat{\chi}\},{\bf S}) +H_{\rm SB}({\bf S})
\label{eq:tot-hamS}
\;.
\end{equation}
The evolution of the density matrix $\hat{\rho}({\bf S})$
of the quantum  system in the classical bath 
can be postulated in the form
\begin{eqnarray}
\frac{\partial}{\partial t}\hat{\rho}({\bf S},t)
&=&-\frac{i}{\hbar}\left[\begin{array}{cc} \hat{\cal H}({\bf S}) & 
\hat{\rho}({\bf S},t)\end{array}\right]
\cdot\mbox{\boldmath$\cal D$}^{\rm S}\cdot
\left[\begin{array}{c}\hat{\cal H}({\bf S})\\ 
\hat{\rho}({\bf S},t)\end{array}\right]
\;,\nonumber\\
&=&-\frac{i}{\hbar}\left[\hat{\cal H}({\bf S}),\hat{\rho}({\bf S},t)
\right]_{\mbox{\tiny \boldmath$\cal D$}^{\rm S}}
\label{eq:pWS}
\end{eqnarray}
where
\begin{eqnarray}
\mbox{\boldmath$\cal D$}^{\rm S}
&=&
\left[\begin{array}{cc} 0 & 1+\frac{i\hbar}{2}
\frac{\overleftarrow{\partial}}{\partial S_I}{\cal B}_{IJ}^{\rm S}
\frac{\overrightarrow{\partial}}{\partial S_J}
\\
-1-\frac{i\hbar}{2}
\frac{\overleftarrow{\partial}}{\partial S_I}{\cal B}_{IJ}^{\rm S}
\frac{\overrightarrow{\partial}}{\partial S_J}
& 0\end{array}\right]\;.
\nonumber\\
\end{eqnarray}
The right-hand side of Eq.~(\ref{eq:pWS}) introduces a quantum-classical
bracket for a quantum subsystem in a classical spin bath.


In order to represent the abstract Eq.~(\ref{eq:pWS}) the
quantum-classical Hamiltonian of Eq.~(\ref{eq:tot-hamS}) can be rewritten as
\begin{equation}
\hat{\cal H}({\bf S}) = \hat{h}({\bf S}) + H_{\rm SB}({\bf S}) \;.
\end{equation}
Accordingly, the adiabatic basis is defined by the eigenvalue equation
\begin{equation}
\hat{h}({\bf S})|\alpha;{\bf S}\rangle =E_{\alpha}({\bf S})|\alpha;{\bf S}\rangle\;.
\label{eq:adbasS}
\end{equation}
However, at variance with the case of the canonical coordinate bath
discussed in Sec.~\ref{sec:qc},
where it depended only on the positions $R$ (and not on the conjugate momenta $P$),
the adiabatic basis defined by Eq.~(\ref{eq:adbasS}) depends on all the non-canonical spin
coordinates $\bf S$.
Hence, in the spin adiabatic basis, Eq.~(\ref{eq:pWS}) becomes
\begin{eqnarray}
\partial_t\rho_{\alpha\alpha'}
&=&
-i\omega_{\alpha\alpha'}\rho_{\alpha\alpha'}
-{\cal B}_{IJ}^{\rm S}\frac{\partial H_{\rm SB}}{\partial S_J}
\langle\alpha;{\bf S}|\frac{\partial\hat{\rho}}{\partial S_I}|\alpha';{\bf S}\rangle
\nonumber\\
&+&\frac{1}{2}
{\cal B}_{IJ}^{\rm S}\langle\alpha;{\bf S}|\frac{\partial\hat{h}}{\partial S_I}
\frac{\partial\hat{\rho}}{\partial S_J}|\alpha';{\bf S}\rangle
\nonumber\\
&-&\frac{1}{2}
{\cal B}_{IJ}^{\rm S}\langle\alpha;{\bf S}|\frac{\partial\hat{\rho}}{\partial S_I}
\frac{\partial\hat{h}}{\partial S_J}|\alpha';{\bf S}\rangle \;,
\end{eqnarray}
where the antisymmetry of $\mbox{\boldmath$\cal B$}^{\rm S}$ has been used.
As it was done in Sec.~\ref{sec:qc}, a coupling vector can be defined as
\begin{eqnarray}
d^I_{\sigma\alpha}&=&\langle\sigma;{\bf S}|
\overrightarrow{\frac{\partial}{\partial S_I}}|\alpha;{\bf S}\rangle \;,
\end{eqnarray}
where the index $I$ of the spin components has been left explicit.
The following identities can be easily found
\begin{eqnarray}
\langle\alpha;{\bf S}|\frac{\partial\hat{\rho}}{\partial S_I}|\alpha';{\bf S}\rangle
&=&
\frac{\partial\rho_{\alpha\alpha'}}{\partial S_I}
+ d^{I}_{\alpha\sigma}\rho_{\sigma\alpha'}
\nonumber\\
&-&\rho_{\alpha\sigma'} d^I_{\sigma'\alpha'} \;,
\label{eq:drho}
\\
\langle\alpha;{\bf S}|\frac{\partial\hat{h}}{\partial S_I}|\sigma;{\bf S}\rangle
&=&
\frac{\partial h_{\alpha\sigma}}{\partial S_I}
-\Delta E_{\alpha\sigma} d^I_{\alpha\sigma} \;,
\label{eq:dh}
\end{eqnarray}
where $\Delta E_{\alpha\sigma}=E_{\alpha}-E_{\sigma}$.
With the help of Eqs.~(\ref{eq:drho}) and~(\ref{eq:dh}),
the equations of motion can be written as
\begin{eqnarray}
\partial_t\rho_{\alpha\alpha'}
&=&-i\omega_{\alpha\alpha'}\rho_{\alpha\alpha'}
- {\cal B}_{IJ}^{\rm S}\frac{\partial H_{\rm SB}}{\partial S_J}
 \frac{\partial\rho_{\alpha\alpha'}}{\partial S_I}
-\frac{1}{2}{\cal B}_{IJ}^{\rm S}\frac{\partial E_{\alpha}}{\partial S_J}
\frac{\partial\rho_{\alpha\alpha'}}{\partial S_I}
\nonumber\\
&-&\frac{1}{2}{\cal B}_{IJ}^{\rm S} \frac{\partial E_{\alpha'}}{\partial S_J}
 \frac{\partial\rho_{\alpha\alpha'}}{\partial S_I}
- \Bigg( -{\cal B}_{IJ}^{\rm S}\frac{\partial H_{\rm SB}}{\partial S_J}
d^{I}_{\alpha\beta}\delta_{\alpha'\beta'}
\nonumber\\
&-&{\cal B}_{IJ}^{\rm S}\frac{\partial H_{\rm SB}}{\partial S_J} 
d^{I*}_{\alpha'\beta'} \delta_{\alpha\beta}
- \frac{1}{2}{\cal B}_{IJ}^{\rm S}\Delta E_{\alpha\beta}
d^{I}_{\alpha\beta}\frac{\partial}{\partial S_J}
\delta_{\alpha'\beta'}
\nonumber\\
&-&\frac{1}{2}{\cal B}_{IJ}^{\rm S}\Delta E_{\alpha'\beta'}
d^{I*}_{\beta'\alpha'}\frac{\partial }{\partial S_J}
\delta_{\alpha\beta}
\Bigg)
\rho_{\beta\beta'}
\nonumber\\
&+&
\frac{1}{2} {\cal B}_{IJ}^{\rm S}
\Bigg(
\frac{\partial E_{\alpha}}{\partial S_I} d^{J}_{\alpha\beta}\delta_{\alpha'\beta'}
-\frac{\partial E_{\alpha}}{\partial S_I} d^J_{\beta'\alpha'}\delta_{\alpha\beta}
\nonumber\\
&-&
\Delta E_{\alpha\sigma}d^{I}_{\alpha\sigma} d^{J}_{\sigma\beta}\rho_{\beta\alpha'}
+\Delta E_{\alpha\beta}d^{I}_{\alpha\beta} d^J_{\beta'\alpha'}
\Bigg)
\rho_{\beta\beta'}
\nonumber\\
&-&\frac{1}{2} {\cal B}_{IJ}^{\rm S}
\Bigg(
\frac{\partial E_{\alpha'}}{\partial S_J} d^{I}_{\alpha\beta}\delta_{\alpha'\beta'}
- \frac{\partial E_{\alpha'}}{\partial S_J} d^I_{\beta'\alpha'} \delta_{\alpha\beta}
\nonumber\\
&+&\Delta E_{\sigma'\alpha'}d^J_{\sigma'\alpha'}
d^I_{\beta'\sigma'} \delta_{\alpha\beta}
- \Delta E_{\beta'\alpha'}d^J_{\beta'\alpha'} d^{I}_{\alpha\beta}
\Bigg)\rho_{\beta\beta'} \;.
\nonumber\\
\end{eqnarray}
At this stage, it is useful to introduce a classical-like spin-Liouville operator:
\begin{eqnarray}
L_{\alpha\alpha'}&=&
\left(
 {\cal B}_{IJ}^{\rm S}\frac{\partial H_{\rm SB}}{\partial S_J} \frac{\partial}{\partial S_I}
+\frac{1}{2} {\cal B}_{IJ}^{\rm S} \frac{\partial E_{\alpha'}}{\partial S_J}
\frac{\partial}{\partial S_I}
+\frac{1}{2} {\cal B}_{IJ}^{\rm S} \frac{\partial E_{\alpha}}{\partial S_J}
\frac{\partial}{\partial S_I}
\right)
\nonumber\\
&=&
{\cal B}_{IJ}^{\rm S}\frac{\partial H_{\alpha\alpha'}^{\rm S}}{\partial S_J}
\frac{\partial}{\partial S_I}
=\left\{...,H_{\alpha\alpha'}^{\rm S}\right\}
_{\mbox{\tiny \boldmath$\cal B$}^{\rm S}}
\;,
\label{eq:Lspin}
\end{eqnarray}
where the matrix elements of the total system Hamiltonian on the adiabatic surfaces 
are denoted as 
\begin{equation}
H_{\alpha\alpha'}^{\rm S}=H_{\rm SB}+\frac{1}{2}\left(E_{\alpha}+E_{\alpha'}\right)
\;.
\end{equation}
A transition operator for the spin bath can be defined as
\begin{eqnarray}
J_{\alpha\alpha',\beta\beta'}
&=&
{\cal B}_{IJ}^{\rm S}\frac{\partial H_{\rm SB}}{\partial S_J}
d^{I}_{\alpha\beta}\delta_{\beta'\alpha'}
+\frac{1}{2}{\cal B}_{IJ}^{\rm S}
\Delta E_{\alpha\beta}d^{I}_{\alpha\beta}\frac{\partial}{\partial S_J}
\delta_{\alpha'\beta'}
\nonumber\\
&+&{\cal B}_{IJ}^{\rm S}\frac{\partial H_{\rm SB}}{\partial S_J}
d^{I*}_{\alpha'\beta'} \delta_{\alpha\beta}
+\frac{1}{2}{\cal B}_{IJ}^{\rm S}\Delta E_{\alpha'\beta'}
d^{I*}_{\alpha'\beta'}\frac{\partial }{\partial S_J}
\delta_{\alpha\beta}
\;.
\nonumber\\
\label{eq:Jspin}
\end{eqnarray}
The operator in Eq.~(\ref{eq:Jspin}) goes to the transition operator
in Eq.(\ref{eq:TO-off})
when canonical variables are considered.
In the case of a spin bath, in order to take properly into account non-adiabatic effects,
a higher order transition operator (acting
together with $J_{\alpha\alpha',\beta\beta'}$) 
must be considered.
Such an operator is
\begin{eqnarray}
{\cal S}_{\alpha\alpha',\beta\beta'}
&=&
\frac{1}{2}{\cal B}_{IJ}^{\rm S}
\frac{\partial (E_{\alpha}+E_{\alpha'})}{\partial S_I} d^{J}_{\alpha\beta}
\delta_{\alpha'\beta'}
\nonumber\\
&+&\frac{1}{2}{\cal B}_{IJ}^{\rm S}
\frac{\partial (E_{\alpha}+E_{\alpha'})}{\partial S_I}
d^{J*}_{\alpha'\beta'}\delta_{\alpha\beta}
\nonumber\\
&-&\frac{1}{2}{\cal B}_{IJ}^{\rm S}\Delta E_{\alpha\sigma}
d^{I}_{\alpha\sigma} d^{J}_{\sigma\beta}\delta_{\beta\alpha'}
-\frac{1}{2}{\cal B}_{IJ}^{\rm S}\Delta E_{\alpha\beta}
d^{I}_{\alpha\beta} d^{J*}_{\alpha'\beta'}
\nonumber\\
&-&\frac{1}{2}{\cal B}_{IJ}^{\rm S}\Delta E_{\alpha'\sigma'}
d^{I*}_{\alpha'\sigma'}d^{J*}_{\sigma'\beta'}\delta_{\alpha\beta}
\nonumber\\
&-&\frac{1}{2}{\cal B}_{IJ}^{\rm S}\Delta E_{\alpha'\beta'}
d^{I*}_{\alpha'\beta'}d^{J}_{\alpha\beta}
\;.
\label{eq:Sspin}
\end{eqnarray}
The operator in Eq.~(\ref{eq:Sspin}) is identically zero for canonical conjugate variables.
Using Eqs.~(\ref{eq:Lspin}-\ref{eq:Sspin}), the equation of motion reads
\begin{eqnarray}
\partial_t\rho_{\alpha\alpha'}
&=&\sum_{\beta\beta'}
\Big(-i\omega_{\alpha\alpha'}\delta_{\alpha\beta}\delta_{\alpha\alpha'}
-L_{\alpha\alpha'}\delta_{\alpha\beta}\delta_{\alpha\alpha'}
-J_{\alpha\alpha',\beta\beta'}
\nonumber\\
&+&{\cal S}_{\alpha\alpha',\beta\beta'}\Big)
\rho_{\beta\beta'} \;.
\label{eq:qcsd-dyna}
\end{eqnarray}

The general equations of motion~(\ref{eq:qcsd-dyna}) are difficult to integrate if
one desires to take into account non-adiabatic corrections.
However, in the case of weak coupling between the spin bath and the quantum subsystem,
one is allowed to take the adiabatic limit
of the operators in Eqs.~(\ref{eq:Jspin}) and~(\ref{eq:Sspin}).
This is performed by assuming that the off-diagonal elements of $d_{\alpha\alpha'}$ (which
couple different adiabatic energy surfaces) are negligible.
In such a case, one obtains
\begin{eqnarray}
J_{\alpha\alpha',\beta\beta'}^{\rm ad}
&=&
-{\cal B}_{IJ}^{\rm S}\frac{\partial H_{\rm SB}}{\partial S_J} 
\left( d^{I}_{\alpha\alpha} + d^{I*}_{\alpha'\alpha'}
\right)\delta_{\alpha\beta}\delta_{\beta'\alpha'}
\nonumber\\
&=&
- i{\cal B}_{IJ}^{\rm S}\frac{\partial H_{\rm SB}}{\partial S_J} 
\left( \Phi^{I}_{\alpha\alpha} - \Phi^{I}_{\alpha'\alpha'}
\right)\delta_{\alpha\beta}\delta_{\beta'\alpha'}
\;,
\nonumber\\
\label{eq:Jspin-ad}
\end{eqnarray}
where, using the fact that $d_{\alpha\alpha}^I$ is purely imaginary, a phase
\begin{equation}
\Phi^I_{\alpha\alpha'}=-id_{\alpha\alpha'}^I
\end{equation}
has been introduced in a manner analogous to that when the bath has
a representation in terms of canonical variables.
In a similar way, one can take the adiabatic limit of the ${\cal S}_{\alpha\alpha',\beta\beta'}$
in Eq.~(\ref{eq:Sspin}):
\begin{eqnarray}
{\cal S}_{\alpha\alpha',\beta\beta'}^{\rm ad}
&=&
-\frac{i}{2}{\cal B}_{IJ}^{\rm S}
\frac{\partial (E_{\alpha}+E_{\alpha'})}{\partial S_J} 
\left(\Phi^{I}_{\alpha\alpha} 
- \Phi^{I}_{\alpha'\alpha'}\right)\delta_{\alpha\alpha}\delta_{\alpha'\alpha'}
\;.\nonumber\\
\label{eq:Sspin-ad}
\end{eqnarray}
Hence, in the weak-coupling (adiabatic) limit, the equation of motion reads
\begin{eqnarray}
\frac{\partial}{\partial t}\rho_{\alpha\alpha'}
&=&\Bigg[-i\omega_{\alpha\alpha'}
-i{\cal B}_{IJ}^{\rm S}\frac{\partial H^{\rm S}_{\alpha\alpha'}}{\partial S_J} 
\left(\Phi^{I}_{\alpha\alpha} - \Phi^{I}_{\alpha'\alpha'}\right)
\nonumber\\
&-&{\cal B}_{IJ}^{\rm S}\frac{\partial H_{\alpha\alpha'}^{\rm S}}{\partial S_J}
\frac{\partial}{\partial S_I} \Bigg]
\rho_{\alpha\alpha'}\;.
\label{eq:qcsd-dyna-ad}
\end{eqnarray}
In the absence of explicit time-dependence in the basis set, Eq.~(\ref{eq:qcsd-dyna-ad})
can be rewritten as
\begin{eqnarray}
\partial_t\rho_{\alpha\alpha'}
&=&\Bigg[-i\omega_{\alpha\alpha'}
-\left( \langle\alpha,{\bf S}|\frac{d}{dt}|\alpha,{\bf S}\rangle -\langle\alpha',{\bf S}|\frac{d}{dt}|\alpha',{\bf S}\rangle \right)
\nonumber\\
&-&{\cal B}_{IJ}^{\rm S}\frac{\partial H_{\alpha\alpha'}^{\rm S}}{\partial S_J}
\frac{\partial}{\partial S_I}
\Bigg] \rho_{\alpha\alpha'} \;.
\label{eq:qcsd-dyna-ad2}
\end{eqnarray}
Using the Dyson identity, this can be written in propagator form as
\begin{eqnarray}
\rho_{\alpha\alpha'}(t)
&=&\exp\left[-i\int_{t_0}^t dt'\omega_{\alpha\alpha'}(t')\right]
\nonumber\\
&\times&\exp\left[
-\int_{t_0}^t dt'\left( \langle\alpha,{\bf S}|\frac{d}{dt'}|\alpha,{\bf S}\rangle
 -\langle\alpha',{\bf S}|\frac{d}{dt'}|\alpha',{\bf S}\rangle \right)\right]
\nonumber\\
&\times&
\exp\left[ -(t-t_0){\cal B}_{IJ}^{\rm S}
\frac{\partial H_{\alpha\alpha'}^{\rm S}}{\partial S_J} 
\frac{\partial}{\partial S_I}
\right]
\rho_{\alpha\alpha'}(t_0)
\;.
\label{eq:qcsd-dyna-ad-prop}
\end{eqnarray}
Equation~(\ref{eq:qcsd-dyna-ad-prop}) provides the adiabatic approximation
of the quantum-classical Liouville equations in spin baths.
The geometric phase arise from the time integral of the
term $\langle\alpha,{\bf S}|(d/dt)|\alpha,{\bf S}\rangle
-\langle\alpha',{\bf S}|(d/dt)|\alpha',{\bf S}\rangle$,
which is purely off-diagonal.


\section{Quantum-classical spin model}
\label{sec:model}

Consider a quantum two-level system represented by the Pauli matrices
\begin{equation}
\sigma_x
=\left[\begin{array}{cc} 0 & 1 \\ 1 & 0\end{array}\right]
\;,\;
\sigma_y
=\left[\begin{array}{cc} 0 & -i \\ i & 0\end{array}\right]
\;,\;
\sigma_z
=\left[\begin{array}{cc} 1 & 0 \\ 0 & -1\end{array}\right]
\;,
\end{equation}
and a classical spin with components ${\bf S}=(S_x,S_y,S_z)$
immersed in a constant magnetic field ${\bf B}=(0,0,b)$.
Hence, consider a model defined by the total Hamiltonian below
\begin{eqnarray}
\hat{H}({\bf S})
&=&
-\Omega \sigma_x - c_1 b \sigma_z -\mu {\bf S}\cdot \mbox{\boldmath $\sigma$} 
- c_2 b S_z + \frac{S_z^2}{2}
\nonumber\\
&=&\hat{h}({\bf S})-c_2bS_z+ \frac{S_z^2}{2} \;,
\label{eq:model-ham}
\end{eqnarray}
where
$\mbox{\boldmath $\sigma$}=(\sigma_x, \sigma_y,\sigma_z)$,
and $c_1, c_2$ are coupling coefficients.

Upon defining
\begin{eqnarray}
\gamma&=& c_1b + \mu S_z \;, \label{eq:gamma}  \\
\eta&=& -\mu S_y  \;,\label{eq:eta} \\
\tilde{\Omega}&=&\Omega+\mu S_x  \;,\label{eq:tilde-omega}
\end{eqnarray}
one can write the eigenvalues of the Hamiltonian~(\ref{eq:model-ham}) as
\begin{eqnarray}
E_1 &=& + \sqrt{\tilde{\Omega}^2+\gamma^2+\eta^2}\;,\label{eq:eigen-E1}
\\
E_2 &=& -\sqrt{\tilde{\Omega}^2+\gamma^2+\eta^2}\;,\label{eq:eigen-E2}
\end{eqnarray}
and the eigenvectors as
\begin{eqnarray}
|E_1\rangle &=& \frac{1}{\sqrt{2(1+|\tilde{G}|^2)}}
\left(\begin{array}{c} 1+\tilde{G}^* \\ \tilde{G}-1\end{array}\right) \;,
\\
|E_2\rangle &=& \frac{1}{\sqrt{2(1+|\tilde{G}|^2)}}
\left(\begin{array}{c} 1-\tilde{G}^* \\ 1+\tilde{G}\end{array}\right) \;,
\end{eqnarray}
where $\tilde{G}=G +i\eta/\gamma$,
and
\begin{eqnarray}
G&=&\frac{-\tilde{\Omega}+\sqrt{\tilde{\Omega}^2+\gamma^2+\eta^2}}{\gamma}\;.
\label{eq:g}
\end{eqnarray}


\subsection{Dynamics on the adiabatic surfaces}

In a Heisenberg-like picture of the dynamics,
quantum-classical operators $\hat{\chi}({\bf S},t))$,
depending on the classical spin coordinates, evolve in time
while the density matrix remain stationary.
From the equation of motion for the density matrix given in~(\ref{eq:qcsd-dyna-ad-prop}),
one can easily obtain the evolution equation for the operator in the adiabatic basis.
For clarity, it us useful to write it explicitly
\begin{eqnarray}
\chi_{\alpha\alpha'}(t)
&=&\exp\left[i\int_{t_0}^t dt'\omega_{\alpha\alpha'}(t')\right]
\nonumber\\
&\times&\exp\left[
\int_{t_0}^t dt'\left( \langle\alpha,S|\frac{d}{dt'}|\alpha,S\rangle
 -\langle\alpha',S|\frac{d}{dt'}|\alpha',S\rangle \right)\right]
\nonumber\\
&\times&
\exp\left[ (t-t_0){\cal B}_{IJ}^{\rm S}
\frac{\partial H_{\alpha\alpha'}^{\rm S}}{\partial S_J}
\frac{\partial}{\partial S_I} \right] \chi_{\alpha\alpha'}(t_0)
\;.
\label{eq:qcsd-op-dyna-ad-prop}
\end{eqnarray}
The equation of motion~(\ref{eq:qcsd-op-dyna-ad-prop}) can be simulated on the
computer in terms of classical-like trajectories evolving on adiabatic energy
surfaces.
The classical-like spin Liouville operator defined in~(\ref{eq:Lspin})
determines the equations of motion 
\begin{eqnarray}
\dot{S}_x&=&S_z\frac{\partial H_{\alpha\alpha'}}{\partial S_y}
-S_y\frac{\partial H_{\alpha\alpha'}}{\partial S_z}
\;,
\label{eq:eqofm1}\\
\dot{S}_y&=&-S_z\frac{\partial H_{\alpha\alpha'}}{\partial S_x}
+S_x\frac{\partial H_{\alpha\alpha'}}{\partial S_z}
\;,
\\
\dot{S}_z&=&S_y\frac{\partial H_{\alpha\alpha'}}{\partial S_x}
-S_x\frac{\partial H_{\alpha\alpha'}}{\partial S_y}
\;,
\label{eq:eqofm3}
\end{eqnarray}
where $H_{\alpha\alpha'}$ are the adiabatic surface Hamiltonians.
For the model in Eq.~(\ref{eq:model-ham}), these can be written explicitly as
\begin{eqnarray}
H_{11}&=&\frac{S_z^2}{2} -c_2 b S_z + \sqrt{\tilde{\Omega}^2+\gamma^2+\eta^2} \;,\\
H_{12}&=&H_{21}= \frac{S_z^2}{2} - c_2 b S_z\;, \\
H_{22}&=&\frac{S_z^2}{2} -c_2 b S_z-\sqrt{\tilde{\Omega}^2+\gamma^2+\eta^2} \;,
\end{eqnarray}
where $\gamma$, $\eta$ and $\tilde{\Omega}$ have been defined in
Eqs.~(\ref{eq:gamma}-\ref{eq:tilde-omega}).

In order to write explicitly the equations of motion for the spin onto
the three energy surfaces,
one needs to calculate the derivatives of the Hamiltonians $H_{11}$, $H_{12}$
and $H_{22}$ with respect to the spin components $(S_x,S_y,S_z)$.
On the $(1,1)$ surface one finds:
\begin{eqnarray}
\frac{\partial H_{11}}{\partial S_x}
&=&
\frac{\mu(\Omega+S_x)}{\sqrt{(\Omega+\mu S_x)^2+\mu^2S_y^2+(\mu S_z -c_1 b)^2}}
\;,\\
\frac{\partial H_{11}}{\partial S_y}
&=&\frac{\mu^2S_y}{\sqrt{(\Omega+\mu S_x)^2+\mu^2S_y^2+(\mu S_z -c_1 b)^2}}
\;,\\
\frac{\partial H_{11}}{\partial S_z}
&=&
S_z-c_2b\nonumber\\
&+&\frac{\mu(b+\mu S_z)}{\sqrt{(\Omega+\mu S_x)^2+\mu^2S_y^2+(\mu S_z -c_1 b)^2}}
\;.
\end{eqnarray}
On the $(1,2)$ and $(2,1)$ surfaces one finds:
\begin{eqnarray}
\frac{\partial H_{12}}{\partial S_x}
&=&\frac{\partial H_{21}}{\partial S_x}=0
\;.\\
\frac{\partial H_{12}}{\partial S_y}
&=&\frac{\partial H_{21}}{\partial S_y}=0
\;.\\
\frac{\partial H_{12}}{\partial S_z}
&=&\frac{\partial H_{21}}{\partial S_z}=S_z - c_2 b
\;,
\end{eqnarray}
while on the $(2,2)$ surface one finds:
\begin{eqnarray}
\frac{\partial H_{22}}{\partial S_x}
&=&
-\frac{\mu(\Omega+S_x)}{\sqrt{(\Omega+\mu S_x)^2+\mu^2S_y^2+(\mu S_z -c_1 b)^2}}
\;,\nonumber\\
\\
\frac{\partial H_{22}}{\partial S_y}
&=&-\frac{\mu^2S_y}{\sqrt{(\Omega+\mu S_x)^2+\mu^2S_y^2+(\mu S_z -c_1 b)^2}}
\;,\nonumber\\
\\
\frac{\partial H_{22}}{\partial S_z}
&=&
S_z-c_2b\nonumber\\
&-&\frac{\mu(b+\mu S_z)}{\sqrt{(\Omega+\mu S_x)^2+\mu^2S_y^2+(\mu S_z -c_1 b)^2}}
\;.
\end{eqnarray}
At this stage, 
in order to simplify the expression of the gradients of the 
adiabatic surface Hamiltonians,
one can use the identity
\begin{eqnarray}
(\Omega+\mu S_x)^2&+&\mu^2S_y^2+(\mu S_z - c_1 b)^2 
=
\Omega^2+c_1^2b^2\nonumber\\
&+&2\mu(\Omega S_x-c_1b S_z)+\mu^2{\bf S}^2 \;.
\end{eqnarray}
As a matter of fact,
the equations of motion~(\ref{eq:eqofm1}-\ref{eq:eqofm3}) conserve the Casimir
${\bf S}^2= S_M S_M$ for an arbitrary Hamiltonian $H_{\alpha\alpha'}$.
To see this, one can rewrite the matrix $\mbox{\boldmath$ \cal B$}^S$
in Eq.~(\ref{eq:bmat}) as
${\cal B}_{IJ}^{\rm S}=\epsilon_{IJK}S_K$ (where $\epsilon_{IJK}$
is the completely anti-symmetric tensor) and obtain
\begin{eqnarray}
\frac{d}{dt}{\bf S}^2
& =&\frac{\partial S_M S_M}{\partial S_I}
{\cal B}_{IJ}^{\rm S}
\frac{\partial H_{\alpha\alpha'}}{\partial S_J}
\nonumber\\
&=&
- 2\epsilon_{MJK}S_MS_K\frac{\partial H_{\alpha\alpha'}}{\partial S_J}
=0\;,
\end{eqnarray}
where in the last step an odd permutations of the indices of
$\epsilon_{KJM}$ has been performed.
Since ${\bf S^2}$ is a constant of motion, one can define
\begin{equation}
C^2=\Omega^2+c_1^2b^2+\mu^2{\bf S}^2\;,
\end{equation}
so that
\begin{eqnarray}
\tilde{\Omega}^2+\gamma^2+\eta^2
&=&C^2+2\mu(\Omega S_x - c_1 b S_z)\;.
\end{eqnarray}
Hence, the derivatives of the adiabatic surfaces can be rewritten as follows.
On the $(1,1)$ surface one has:
\begin{eqnarray}
\frac{\partial H_{11}}{\partial S_x}
&=&
\frac{\mu(\Omega+S_x)}{\sqrt{C^2+2\mu(\Omega S_x -c_1 b S_z)}}
\;,\\
\frac{\partial H_{11}}{\partial S_y}
&=&\frac{\mu^2S_y}{\sqrt{C^2+2\mu(\Omega S_x -c_1 b S_z)}}
\;,\\
\frac{\partial H_{11}}{\partial S_z}
&=&
S_z-c_2b\nonumber\\
&+&\frac{\mu(b+\mu S_z)}{\sqrt{C^2+2\mu(\Omega S_x -c_1 b S_z)}}
\;.
\end{eqnarray}
On the $(1,2)$ and $(2,1)$ surfaces one has:
\begin{eqnarray}
\frac{\partial H_{12}}{\partial S_x}
&=&\frac{\partial H_{21}}{\partial S_x}=0
\;,\\
\frac{\partial H_{12}}{\partial S_y}
&=&\frac{\partial H_{21}}{\partial S_y}=0
\;,\\
\frac{\partial H_{12}}{\partial S_z}
&=&\frac{\partial H_{21}}{\partial S_z}= S_z - c_2 b
\;,
\end{eqnarray}
and on the $(2,2)$ surface one has:
\begin{eqnarray}
\frac{\partial H_{22}}{\partial S_x}
&=&
-\frac{\mu(\Omega+S_x)}{\sqrt{C^2+2\mu(\Omega S_x - c_1 b S_z)}}
\;,\\
\frac{\partial H_{22}}{\partial S_y}
&=&-\frac{\mu^2S_y}{\sqrt{C^2+2\mu(\Omega S_x - c_1 b S_z)}}
\;,\\
\frac{\partial H_{22}}{\partial S_z}
&=&
S_z-c_2b\nonumber\\
&-&\frac{\mu(b+\mu S_z)}{\sqrt{C^2+2\mu(\Omega S_x - c_1 b S_z)}}
\;.
\end{eqnarray}


\section{Time-reversible integrators}
\label{sec:integrator}

A different set of equations of motion corresponds to each adiabatic energy surface.
Hence, one has to find different algorithms of integration on each surface.
In the following, the Liouville propagator on each surface is factorized
and the associated time-reversible algorithm for the spin dynamics is derived.
It is worth noting that within a purely classical context other authors have considered
alternative schemes of integration~\cite{leimkuhler,landau,steinigeweg}.
At the same time,
while what follows is based on the basic symmetric Trotter factorization of
the evolution operator, 
in order to integrate the spin dynamics,
one could have used the elegant time-reversible measure-preserving
algorithms invented by G. S. Ezra~\cite{ezra,sergi-ezra,sergi-ezra2}.

\subsection{Reversible integrator on the $(1,1)$ adiabatic surface}

The equations of motion on the $(1,1)$ surface can be written explicitly as
\begin{eqnarray}
\dot{S}_x
&=&
\frac{\mu^2S_yS_z-\mu S_y(b+\mu S_z)}{\sqrt{C^2+2\mu(\Omega S_x - c_1 b S_z)}}
\nonumber\\
&-&S_y\left(S_z-c_2b\right)\;,
\label{eq:dotsx11}
\\
\dot{S}_y
&=&
\frac{\mu S_x(b+\mu S_z)-\mu(\Omega+S_x)S_z}{\sqrt{C^2+2\mu(\Omega S_x - c_1 b S_z)}}
\nonumber\\
&+&S_x\left(S_z-c_2b\right)\;,
\label{eq:dotsy11}
\\
\dot{S}_z
&=&\frac{\mu(\Omega+S_x)S_y-\mu^2S_xS_y}{\sqrt{C^2+2\mu(\Omega S_x - c_1 b S_z)}} \;.
\label{eq:dotsz11}
\end{eqnarray}
From Equations~(\ref{eq:dotsx11}-\ref{eq:dotsz11}) one can easily find
the corresponding Liouville operators
\begin{eqnarray}
L_{1,(1,1)}^{S_x}&=&
\frac{\mu^2S_yS_z-\mu S_y(b+\mu S_z)}
{\sqrt{C^2+2\mu(\Omega S_x - c_1 b S_z)}}\frac{\partial}{\partial S_x} \;,
\label{eq:L111}
\\
L_{2,(1,1)}^{S_x}&=&-S_y\left(S_z-c_2b\right)\frac{\partial}{\partial S_x} \;,
\\
L_{(1,1)}^{S_y}&=&\Bigg[
\frac{\mu S_x(b+\mu S_z)-\mu(\Omega+S_x)S_z}{\sqrt{C^2+2\mu(\Omega S_x - c_1 b S_z)}}
\nonumber\\
&+&S_x\left(S_z-c_2b\right)
\Bigg]\frac{\partial}{\partial S_y} \;,
\\
L_{(1,1)}^{S_z}&=&
\frac{\mu(\Omega+S_x)S_y-\mu^2S_xS_y}{\sqrt{C^2+2\mu(\Omega S_x - c_1 b S_z)}}
\frac{\partial}{\partial S_z} \;,
\end{eqnarray}
and the propagators
\begin{eqnarray}
U_{1,(1,1)}^{S_x}(\tau)&=&\exp\left[\tau
\frac{\mu^2S_yS_z-\mu S_y(b+\mu S_z)}
{\sqrt{C^2+2\mu(\Omega S_x - c_1 b S_z)}}\frac{\partial}{\partial S_x} \right] \;,
\nonumber\\
\\
U_{2,(1,1)}^{S_x}(\tau)&=&\exp\left[-\tau S_y\left(S_z-c_2b\right)
\frac{\partial}{\partial S_x}\right] \;,
\\
U_{(1,1)}^{S_y}(\tau)&=&\exp\Bigg\{\tau\Bigg[
\frac{\mu S_x(b+\mu S_z)-\mu(\Omega+S_x)S_z}{\sqrt{C^2+2\mu(\Omega S_x - c_1 b S_z)}}
\nonumber\\
&+&S_x\left(S_z-c_2b\right) \Bigg]\frac{\partial}{\partial S_y}\Bigg\} \;,
\\
U_{(1,1)}^{S_z}(\tau)&=&\exp\left[\tau
\frac{\mu(\Omega+S_x)S_y-\mu^2S_xS_y}{\sqrt{C^2+2\mu(\Omega S_x - c_1 b S_z)}}
\frac{\partial}{\partial S_z}\right] \;.
\nonumber\\
\end{eqnarray}

The action of $U_{1,(1,1)}^{S_x}(\tau)=\exp[\tau L_{1,(1,1)}^{S_x}]$
on $S_x$ can be found by means of the analytical integration
of the equation of motion associated to the Liouville operator
$L_{1,(1,1)}^{S_x}$ in Eq.~(\ref{eq:L111}).
In pseudo-code form, one obtains
\begin{eqnarray}
U_{1,(1,1)}^{S_x}(\tau) &:& 
\left\{
\begin{array}{lcl}
S_x &\to& \frac{1}{C_1}\left\{\frac{3}{2}C_1C_3\tau +[C_2+C_1S_x(0)]^{\frac{3}{2}}
\right\}^{\frac{2}{3}}\\
&-&\frac{C_2}{C_1}  \;,
\end{array}
\right. 
\nonumber\\
\end{eqnarray}
where
\begin{eqnarray}
C_1&=&2\mu\Omega \label{eq:defC_1} \;, \\
C_2&=&C^2 - 2\mu c_1 b S_z\label{eq:defC_2} \;, \\
C_3&=&\mu^2S_yS_z-\mu S_y(b+\mu S_z) \label{eq:defC_3} \;.
\end{eqnarray}
The actions of $U_{2,(1,1)}^{S_x}(\tau)=\exp[\tau L_{2,(1,1)}^{S_x}]$ and
$U_{(1,1)}^{S_y}(\tau)=\exp[\tau L_{(1,1)}^{S_y}]$
on $S_x$ and $S_y$, respectively, are simple variable shifts which can be
written in pseudo-code form as
\begin{eqnarray}
U_{2,(1,1)}^{S_x}(\tau)&:& \Big\{
S_x \to S_x - \tau S_y\left(S_z-c_2 b\right)  \;,
\\
U_{(1,1)}^{S_y}(\tau)&:& 
\left\{
\begin{array}{lcl}
S_y &\to& S_y + 
\tau\Big[
\frac{\mu S_x(b+\mu S_z)-\mu(\Omega+S_x)S_z}{\sqrt{C^2+2\mu(\Omega S_x - c_1 b S_z)}}
\\
&+&S_x\left(S_z-c_2b\right) \Big]  \;.
\end{array}
\right.
\nonumber\\
\end{eqnarray}
The action of $U_{(1,1)}^{S_z}(\tau)=\exp[\tau L_{(1,1)}^{S_z}]$
on $S_z$ can also be determined by the analytical integration
of the corresponding equation of motion
\begin{eqnarray}
\dot{S}_z
&=&
\frac{\mu(\Omega+S_x)S_y-\mu^2S_xS_y}{\sqrt{C^2+2\mu(\Omega S_x - c_1 b S_z)}}
=\frac{B_3}{\sqrt{B_2 - B_1S_z}} \;,\nonumber\\
\end{eqnarray}
where one has defined
\begin{eqnarray}
B_1&=&2\mu c_1 b \label{eq:defB_1} \;, \\
B_2&=&C^2+2\mu \Omega S_x \label{eq:defB_2} \;, \\
B_3&=&\mu(\Omega+S_x)S_y-\mu^2S_xS_y \;. \label{eq:defB_3}
\end{eqnarray}
One finds
\begin{eqnarray}
U_{(1,1)}^{S_z}(\tau)
&:&
\left\{
\begin{array}{lcl}
 S_z& \to & \frac{B_2}{B_1} \\
&-&\frac{1}{B_1}\left\{\left[B_2 -
B_1S_z\right]^{\frac{3}{2}}-\frac{3 B_1 B_3\tau}{2}\right\}^{\frac{2}{3}} \;.
\end{array}
\right.
\nonumber\\
\end{eqnarray}

Finally, one can consider three propagators on the $(1,1)$ surface:
\begin{eqnarray}
U_{(1,1)}^1(\tau)
&=&
U_{1,(1,1)}^{S_x}\left(\frac{\tau}{4}\right)
U_{2,(1,1)}^{S_x}\left(\frac{\tau}{2}\right)
U_{1,(1,1)}^{S_x}\left(\frac{\tau}{4}\right)
\nonumber\\
&\times&
U_{(1,1)}^{S_y}\left(\frac{\tau}{2}\right)
U_{(1,1)}^{S_z}(\tau)
U_{(1,1)}^{S_y}\left(\frac{\tau}{2}\right)
\nonumber\\
&\times&
U_{1,(1,1)}^{S_x}\left(\frac{\tau}{4}\right) 
U_{2,(1,1)}^{S_x}\left(\frac{\tau}{2}\right)
U_{1,(1,1)}^{S_x}\left(\frac{\tau}{4}\right) \;,
\nonumber\\
\\
U_{(1,1)}^2(\tau)
&=&
U_{(1,1)}^{S_z}\left(\frac{\tau}{2}\right)
U_{1,(1,1)}^{S_x}\left(\frac{\tau}{4}\right)
U_{2,(1,1)}^{S_x}\left(\frac{\tau}{2}\right)
\nonumber\\
&\times&
U_{1,(1,1)}^{S_x}\left(\frac{\tau}{4}\right)
U_{(1,1)}^{S_y}(\tau)
U_{1,(1,1)}^{S_x}\left(\frac{\tau}{4}\right)
\nonumber\\
&\times&
U_{2,(1,1)}^{S_x}\left(\frac{\tau}{2}\right)
U_{1,(1,1)}^{S_x}\left(\frac{\tau}{4}\right)
U_{(1,1)}^{S_z}\left(\frac{\tau}{2}\right) \;,
\nonumber \\
\\
U_{(1,1)}^3(\tau)
&=&
U_{(1,1)}^{S_y}\left(\frac{\tau}{4}\right)
U_{(1,1)}^{S_z}\left(\frac{\tau}{2}\right)
U_{(1,1)}^{S_y}\left(\frac{\tau}{4}\right)
\nonumber\\
&\times&
U_{1,(1,1)}^{S_x}\left(\frac{\tau}{2}\right)
U_{2,(1,1)}^{S_x}(\tau)
U_{1,(1,1)}^{S_x}\left(\frac{\tau}{2}\right)
\nonumber\\
&\times&
U_{(1,1)}^{S_y}\left(\frac{\tau}{4}\right)
U_{(1,1)}^{S_z}\left(\frac{\tau}{2}\right)
U_{(1,1)}^{S_y}\left(\frac{\tau}{4}\right) \;,
\end{eqnarray}
and write the corresponding integration algorithm.
Each of the three propagators $U_{(1,1)}^k(\tau)$, $k=1,...,3$, can be
used to obtain a different propagation algorithm.
In order to obtain a more uniform sampling of phase space, one can also
act with a different $U_{(1,1)}^k(\tau)$ at each successive time step $\tau$.


\subsection{Reversible integrator on the $(2,2)$ adiabatic surface}

The equations of motion on the $(2,2)$ surface are
\begin{eqnarray}
\dot{S}_x&=&
\frac{\mu S_y(b+\mu S_z)-\mu^2S_yS_z}{\sqrt{C^2+2\mu(\Omega S_x - c_1 b S_z)}}
\nonumber\\
&-&S_y\left(S_z-c_2b\right) \;,
\label{eq:dotsx22}
\\
\dot{S}_y&=&
\frac{\mu(\Omega+S_x)S_z-\mu S_x(b+\mu S_z)}{\sqrt{C^2+2\mu(\Omega S_x - c_1 b S_z)}}
\nonumber\\
&+&S_x\left(S_z-c_2b\right) \;,
\\
\dot{S}_z&=&\frac{\mu^2S_xS_y-\mu(\Omega+S_x)S_y}{\sqrt{C^2+2\mu(\Omega S_x - c_1 b S_z)}}
\;.
\label{eq:dotsz22}
\end{eqnarray}
From Equations~(\ref{eq:dotsx22}-\ref{eq:dotsz22}) one can easily write
the corresponding Liouville operators
\begin{eqnarray}
L_{1,(2,2)}^{S_x}&=&
-\frac{C_3}{\sqrt{C_2+C_1S_x}} \frac{\partial}{\partial S_x} \;,
\label{eq:L1sx22}
\\
L_{2,(2,2)}^{S_x}&=&-S_y\left(S_z-c_2b\right)\frac{\partial}{\partial S_x}
\;,
\label{eq:L2sx22}
\\
L_{(2,2)}^{S_y}&=&\Bigg[
\frac{\mu(\Omega+S_x)S_z-\mu S_x(b+\mu S_z)}{\sqrt{C^2+2\mu(\Omega S_x - c_1 b S_z)}}
\nonumber\\
&+&S_x\left(S_z-c_2b\right) \Bigg]\frac{\partial}{\partial S_y} \;,
\label{eq:Lsy22}
\\
L^{S_z}_{(2,2)}&=&
=-\frac{B_3}{\sqrt{B_2-B_1S_z}}\frac{\partial}{\partial S_z} \;,
\label{eq:Lsz22}
\end{eqnarray}
where $C_1,C_2,C_3$ have been defined in Eqs.~(\ref{eq:defC_1}-\ref{eq:defC_3})
and $B_1,B_2,B_3$ have been defined in Eqs.~(\ref{eq:defB_1}-\ref{eq:defB_3}).
The propagators for the $(2,2)$ adiabatic surface can be written as
\begin{eqnarray}
U_{1,(2,2)}^{S_x}(\tau)&=&\exp\left[
-\tau\frac{C_3}{\sqrt{C_2+C_1S_x}}
\frac{\partial}{\partial S_x}\right] \;,
\\
U_{2,(2,2)}^{S_x}(\tau)&=&\exp\left[-\tau S_y\left(S_z-c_2b\right)
\frac{\partial}{\partial S_x}\right] \;,
\\
U^{S_y}_{(2,2)}(\tau)&=&\exp\Bigg\{\tau\left[
\frac{\mu(\Omega+S_x)S_z-\mu S_x(b+\mu S_z)}{\sqrt{C^2+2\mu(\Omega S_x - c_1 b S_z)}}
\right.\nonumber\\
&+&\left.S_x\left(S_z-c_2b\right)
\right]\frac{\partial}{\partial S_y}\Bigg\} \;,
\\
U^{S_z}_{(2,2)}(\tau)&=&
\exp\left[-\tau\frac{B_3}{\sqrt{B_2-B_1S_z}}\frac{\partial}{\partial S_z}\right] \;,
\end{eqnarray}
where $C_1,C_2,C_3$ have been defined in Eqs.~(\ref{eq:defC_1}-\ref{eq:defC_3})
and $B_1,B_2,B_3$ have been defined in Eqs.~(\ref{eq:defB_1}-\ref{eq:defB_3}).

The action of $U_{1,(2,2)}^{S_x}(\tau)=\exp[L_{1,(2,2)}^{S_x}]$ on $S_x$ is
determined by integrating analytically the equation of motion
associated to the Liouville operator in~(\ref{eq:L1sx22}).
In pseudo-code form, such action can be written as
\begin{eqnarray}
U_{1,(2,2)}^{S_x}(\tau)
&:&\left\{
\begin{array}{lcl}
S_x &\to &
\frac{1}{C_1}\left[-\frac{3}{2}C_1C_3\tau +\left(C_2+C_1S_x\right)^{3/2}
\right]^{2/3}\\
&-&\frac{C_2}{C_1}\;.
\end{array}
\right.
\nonumber\\
\end{eqnarray}
Similarly, the action of $U_{(2,2)}^{S_z}(\tau)=\exp[L_{(2,2)}^{S_z}]$ on $S_z$ 
is determined by integrating analytically the equation of motion associated
to the Liouville operator in~(\ref{eq:Lsz22}): 
\begin{eqnarray}
U_{(2,2)}^{S_z}(\tau)
&:&\left\{
S_z \to
\frac{B_2}{B_1} -\frac{1}{B_1}\left[\left(B_2-B_1S_z\right)^{\frac{3}{2}}
+ \frac{3}{2} B_1B_3 \tau\right]^{\frac{2}{3}} \right. \;.
\nonumber\\
\end{eqnarray}
The propagators $U^{S_y}_{(2,2)}(\tau)$ and $U^{S_z}_{(2,2)}(\tau)$ generate
simple time-shifts of the appropriate spin coordinates:
\begin{eqnarray}
U_{2,(2,2)}^{S_x}(\tau)
&:& \Big\{
S_x \to S_x -\tau S_y\left(S_z-c_2b\right)
\;.
\\
U_{(2,2)}^{S_y}(\tau)
&:& \left\{
\begin{array}{lcl}
S_y &\to & S_y + \tau\Bigg[
\frac{\mu(\Omega+S_x)S_z-\mu S_x(b+\mu S_z)}{\sqrt{C^2+2\mu(\Omega S_x - c_1 b S_z)}}
\\
&+&S_x\left(S_z-c_2b\right) \Bigg]\;.
\end{array}
\right. 
\nonumber\\
\end{eqnarray}

Finally, one can consider the following three propagators on the $(2,2)$ surface:
\begin{eqnarray}
U_{(2,2)}^1(\tau)&=&
U_{1,(2,2)}^{S_x}\left(\frac{\tau}{4}\right)
U_{2,(2,2)}^{S_x}\left(\frac{\tau}{2}\right)
U_{1,(2,2)}^{S_x}\left(\frac{\tau}{4}\right)
\nonumber\\
&\times&
U_{(2,2)}^{S_y}\left(\frac{\tau}{2}\right)
U_{(2,2)}^{S_z}(\tau)
U_{(2,2)}^{S_y}\left(\frac{\tau}{2}\right)
\nonumber\\
&\times&
U_{1,(2,2)}^{S_x}\left(\frac{\tau}{4}\right)
U_{2,(2,2)}^{S_x}\left(\frac{\tau}{2}\right)
U_{1,(2,2)}^{S_x}\left(\frac{\tau}{4}\right) \;,
\nonumber\\
\\
U_{(2,2)}^2(\tau)&=&
U_{(2,2)}^{S_z}\left(\frac{\tau}{2}\right)
U_{1,(2,2)}^{S_x}\left(\frac{\tau}{4}\right)
U_{2,(2,2)}^{S_x}\left(\frac{\tau}{2}\right)
\nonumber\\
&\times&
U_{1,(2,2)}^{S_x}\left(\frac{\tau}{4}\right)
U_{(2,2)}^{S_y}(\tau)
U_{1,(2,2)}^{S_x}\left(\frac{\tau}{4}\right)
\nonumber\\
&\times&
U_{2,(2,2)}^{S_x}\left(\frac{\tau}{2}\right)
U_{1,(2,2)}^{S_x}\left(\frac{\tau}{4}\right)
U_{(2,2)}^{S_z}\left(\frac{\tau}{2}\right) \;,
\nonumber\\
\\
U_{(2,2)}^3(\tau)&=&
U_{(2,2)}^{S_y}\left(\frac{\tau}{4}\right)
U_{(2,2)}^{S_z}\left(\frac{\tau}{2}\right)
U_{(2,2)}^{S_y}\left(\frac{\tau}{4}\right)
\nonumber\\
&\times&
U_{1,(2,2)}^{S_x}\left(\frac{\tau}{2}\right)
U_{2,(2,2)}^{S_x}(\tau)
U_{1,(2,2)}^{S_x}\left(\frac{\tau}{2}\right)
\nonumber\\
&\times&
U_{(2,2)}^{S_y}\left(\frac{\tau}{4}\right)
U_{(2,2)}^{S_z}\left(\frac{\tau}{2}\right)
U_{(2,2)}^{S_y}\left(\frac{\tau}{4}\right) \;,
\nonumber\\
\end{eqnarray}
These allow one to find the algorithm of propagation
on the $(2,2)$ surface.
Each of the three propagators $U_{(2,2)}^k(\tau)$, $k=1,...,3$, can be
used to obtain a different propagation algorithm.
In order to obtain a more uniform sampling of phase space, one can also
act with a different $U_{(2,2)}^k(\tau)$ at each successive time step $\tau$.


\subsection{Reversible integrators on the $(1,2)$ adiabatic surface}

The equations of motion on the $(1,2)$ adiabatic energy surface are
\begin{eqnarray}
\dot{S}_x&=&-S_y\left(S_z-c_2b\right) \;,
\label{eq:dotsx12} \\
\dot{S}_y&=&S_x\left(S_z-c_2b\right)\;,
\label{eq:dotsy12}\\
\dot{S}_z&=&0\;. \label{eq:dotsz12}
\end{eqnarray}
They are identical to the equations of motion on the $(2,1)$ energy surface.
The Liouville operators associated to the Eqs.~(\ref{eq:dotsx12}) and~(\ref{eq:dotsy12})
are:
\begin{eqnarray}
L_{(1,2)}^{S_x}&=&-S_y\left(S_z-c_2b\right)\frac{\partial}{\partial S_x} \;,
\label{eq:Lsx12}
\\
L_{(1,2)}^{S_y}&=&S_x\left(S_z-c_2\right)\frac{\partial}{\partial S_y} \;.
\label{eq:Lsy12}
\end{eqnarray}
The associated propagators are:
\begin{eqnarray}
U_{(1,2)}^{S_x}(\tau)&=&\exp\left[-\tau S_y\left(S_z-c_2b\right)
\frac{\partial}{\partial S_x}\right]\;,
\\
U_{(1,2)}^{S_y}(\tau)&=&\exp\left[\tau S_x\left(S_z-c_2b\right)
\frac{\partial}{\partial S_y}\right]\;.
\end{eqnarray}

Finally, one can consider the following total propagators on the $(1,2)$ surface:
\begin{eqnarray}
U_{(1,2)}^1(\tau)&=&U_{(1,2)}^{S_x}\left(\frac{\tau}{2}\right)
U_{(1,2)}^{S_y}(\tau)
U_{(1,2)}^{S_x}\left(\frac{\tau}{2}\right) \;,
\\
U_{(1,2)}^2(\tau)&=& 
U_{(1,2)}^{S_y}\left(\frac{\tau}{2}\right)
U_{(1,2)}^{S_x}(\tau) 
U_{(1,2)}^{S_y}\left(\frac{\tau}{2}\right)\;,
\end{eqnarray}
and easily write the algorithm of integration.
Each of the two propagators $U_{(1,2)}^k(\tau)$, $k=1,2$, can be
used to obtain a different propagation algorithm.
In order to obtain a more uniform sampling of phase space, one can also
act with a different $U_{(1,2)}^k(\tau)$ at each successive time step $\tau$.


\section{Numerical results}
\label{sec:results}

In order to analyze the quantum dynamics of the model, one can calculate
averages in the Heisenberg-like picture:
\begin{eqnarray}
\langle\hat{\chi}\rangle_t
&=&\sum_{\alpha\alpha'}\int d^3{\bf S} \rho_{\alpha\alpha'}({\bf S})
\chi_{\alpha'\alpha}({\bf S},t)
\label{eq:chi_t}
\end{eqnarray}
where $\hat{\chi}({\bf S},t)$ is the chosen observable (which is evolved in time),
also depending
on the classical spin coordinates ${\bf S}=(S_x,S_y,S_z)$,
and $d^3{\bf S}=dS_xdS_ydS_z$.
For the sake of illustrating the integration algorithms derived in 
Sec.~\ref{sec:integrator},
it is assumed that at time $t=0$ the spin and the quantum systems are decoupled so
that the initial density matrix is
\begin{equation}
\hat{\rho}({\bf S})=\hat{\rho}_{\rm s}\otimes
\sqrt{\frac{\beta}{2\pi}}\exp[-\beta S_z^2] \;,
\end{equation}
where $\hat{\rho}_{\rm s}$ is the density matrix of the isolated quantum subsystem.
In order to study the evolution of both the population difference
between the excited and ground state of the model and the coherence of
the initial superposition between such states,
it is assumed that the quantum subsystem is in a superposition of states at $t=0$
that is represented in the basis of $\sigma_z$ by the state vector
\begin{equation}
|\Psi\rangle = \frac{\sqrt{5}}{5}\left[2|1\rangle - |2\rangle\right]
\;.
\end{equation}
The associated density matrix has components
\begin{equation}
\mbox{\boldmath$\rho$}_{\rm s}=
\left[\begin{array}{cc} 4/5 & -2/5 \\ -2/5 & 1/5 \end{array}\right]
\;.
\end{equation}
According to such choices, the dynamics will display both population
and coherence oscillations.
\begin{figure}
\resizebox{7cm}{5cm}{
\includegraphics* {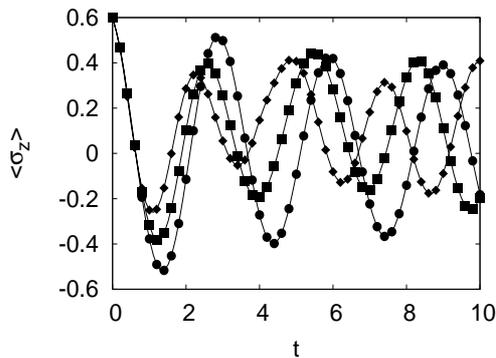}
}
\caption{Time evolution of the population difference $\langle\sigma_z(t)\rangle$
 for $\beta=0.3$,
$\Omega=1$, $b=1$, $c_1=0.01$, $c_2=0.1$. The coefficient $\mu$
takes the values 0.25, 0.5 and 0.74 for black circles, squares and
diamonds, respectively. The lines are drawn to help the eye. 
}
\label{fig:fig1}
\end{figure}
The density matrix in the adiabatic basis takes the form
\begin{eqnarray}
\mbox{\boldmath$\rho$}_{\rm s}^{\rm ad}
&=&
\frac{1}{\cal N}
\left[\begin{array}{cc}
\rho_{11} & \rho_{12}
\\
\rho_{12}^* &
\rho_{22}
\end{array}\right]
\;,
\nonumber\\
\end{eqnarray}
where, using the definition of $G$ given in Eq.~(\ref{eq:g}), 
\begin{eqnarray}
\rho_{11}&=&\frac{1}{5}\left[\frac{9\eta^2}{\gamma^2}+(3+G)^2\right] \;,
\\
\rho_{22}&=&\frac{1}{5}\left[\frac{\eta^2}{\gamma^2}+(1-G)^2\right]\;,
\\
\rho_{12}&=&-\frac{[3i\eta+\gamma(3+G)][-i\eta+\gamma(-1+3G)]}{5\gamma^2}\;,
\nonumber\\
\\
{\cal N}&=&2(1+G^2+\eta^2/\gamma^2)\;.
\end{eqnarray}

\begin{figure}
\resizebox{7cm}{5cm}{
\includegraphics* {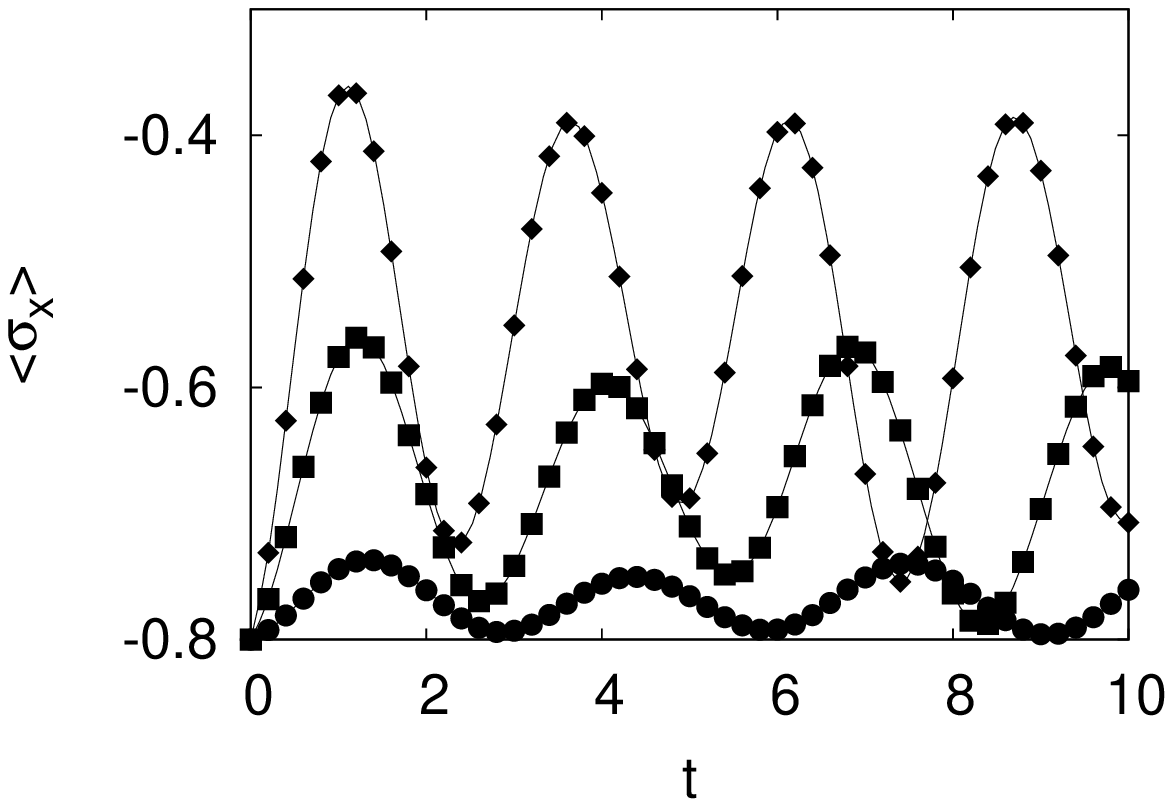}
}
\caption{
Time evolution of the coherence $\langle\sigma_x(t)\rangle$ for $\beta=0.3$,
$\Omega=1$, $b=1$, $c_1=0.01$, $c_2=0.1$. The coefficient $\mu$
takes the values 0.25, 0.5 and 0.74 for black circles, squares and
diamonds, respectively. The lines are drawn to help the eye.
}
\label{fig:fig2}
\end{figure}
One can use spherical coordinates
\begin{eqnarray}
S_x & = &  S \sin(\theta) \cos(\phi)\\
S_y & = &  S \sin(\phi) \sin(\theta) \\
S_z & = &  S \cos(\theta)
\end{eqnarray}
in order to sample the Boltzmann weight on on $S_z$ as
\begin{equation}
\exp\left[-\beta\frac{S_z^2}{2}\right]
=\exp\left[-\beta\frac{\cos^2\theta}{2}\right]
\end{equation}
by sampling $cos(\theta)$ uniformly between $(-1,1)$
and to sample the angle $\phi$ uniformly between $(0,2\pi)$.
\begin{figure}
\resizebox{7cm}{5cm}{
\includegraphics* {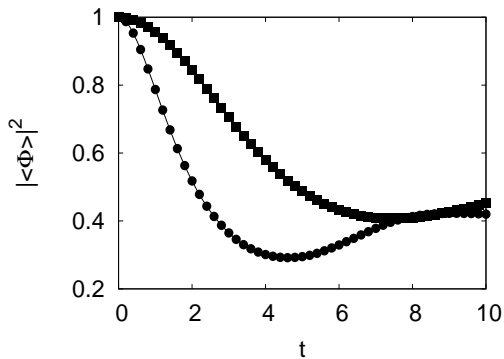}
}
\caption{Time evolution of the modulus square of the average phase
for $\beta=0.3$,
$\Omega=1$, $b=1$, $c_1=0.01$, $c_2=0.1$, $\mu=0.25$;
the black circles denote the results for the geometric phase while the black
squares denote the results for the Bohr phase.
}
\label{fig:fig3}
\end{figure}
The observables $\sigma_z({\bf S},t)$ and $\sigma_x({\bf S},t)$ are evolved
in the adiabatic basis and the calculation of their trace, according to
Eq.~(\ref{eq:chi_t}), provides the population and the coherence evolution, respectively.

Calculations were performed 
for $\beta=0.3$, $\Omega=1$, $b=1$, $c_1=0.01$, $c_2=0.1$. The coefficient $\mu$
was varied and took the values 0.25, 0.5 and 0.74.
The values of the parameters are given in dimensionless units.
Figure~\ref{fig:fig1} shows the behavior of the population as a function of time
when the coupling $\mu$ is varied. The damping increases as the coupling increases.
Figure~\ref{fig:fig2} displays the time evolution of the coherence when the coupling
is varied. The coherence oscillations are greater for greater coupling.
Since the two-level system is coupled to a single rotating classical spin,
no real dissipation is expected when monitoring the dynamics of the two-level
system only.
In Figure~\ref{fig:fig3} the time evolution of the moduli square of the average
of the Bohr and geometric phases are shown. No major geometric effect was expected
for the model studied.
Finally, the stability of the integration algorithm introduced in 
Sec.~(\ref{sec:integrator}) is illustrated in Fig.~(\ref{fig:fig4}).
A numerical integration time step $\tau= 0.001$ (in dimensionless units)
was used in all the calculation performed. The Trotter symmetric factorization
discussed in Sec.~(\ref{sec:integrator}) was combined with a fifth order
Yoshida scheme~\cite{yoshida}.
As expected,
the more stable numerical integration is achieved on the $(1,2)$ mean energy surface.
This arises from the absence of quantum effects on the mean surface of the model in 
Eq.~(\ref{eq:model-ham}).
As quantum effects increase, going from the ground state $(2,2)$ to
the excited state $(1,1)$, the stability of the numerical integration
somewhat diminishes but remains satisfactory
over the whole time interval explored.
The numerical conservation of the spin modulus is almost perfect on all
the three adiabatic energy surfaces.

\begin{figure}
\resizebox{7cm}{5cm}{
\includegraphics* {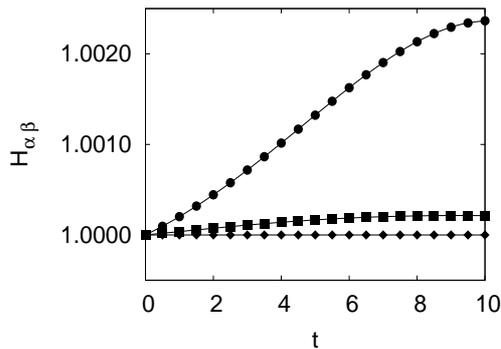}
}
\caption{Adiabatic surface Hamiltonians vs time.
The black circles denote the curve for $\alpha=1$ and $\beta=1$
(excited state dynamics); the square denote the curve for $\alpha=2$ and $\beta=2$
(ground state dynamics) while the black diamonds denote the curve
$\alpha=1$ and $\beta=2$ (mean surface dynamics).
The continues line are for helping the eye.
The parameters specifying the calculations are
$\beta=0.3$, $\Omega=1$, $b=1$, $c_1=0.01$, $c_2=0.1$, $\mu=0.75$.
the black circles denote the results for the geometric phase while the black
}
\label{fig:fig4}
\end{figure}

\section{Conclusions and perspectives}
\label{sec:conc}

In this paper a formalism for studying the dynamics of quantum systems 
coupled to classical spin environments has been reviewed.
The theory is based on generalized antisymmetric brackets
and naturally predicts the existence of open-path off-diagonal
geometric phases in the dynamics of the density matrix.
It has also been shown that such geometric phases must be considered
in the quantum-classical Liouville equation, expressed by means
of canonical phase space coordinates, whenever the adiabatics basis
is complex (as in the case of a magnetic field coupled to the quantum subsystem).

When the quantum subsystem is weakly coupled to the spin environment,
non-adiabatic transitions can be neglected.
In such a case, one can construct an effective non-Markovian computer simulation scheme
for open quantum system dynamics in a classical spin environment. 
In this paper a detailed derivation of
integration algorithms based on the symmetric Trotter factorization
of the classical-like spin propagator has been given.
Such algorithms have been applied to a model system comprising 
a quantum two-level system
coupled to a single classical spin in an external magnetic field.
The numerical integration conserves the spin modulus perfectly
and the spin energy satisfactorily during the entire time interval explored.
Starting from an excited state, the population difference and the
coherences of the two-state model have been simulated and studied
in function of the strength of the coupling parameter
between the spin and the two-level system.

One could look at
the numerical evidence provided in this paper
as a first step toward developing the simulation schemes for quantum dynamics
in classical spin environments into an effective tool for studying
systems of interest in nano-science.

\appendix

\section{Integration algorithm on the $(1,1)$ surface}
\label{app:int-11}

In pseudo-code form, the algorithm provided by $U_{(1,1)}^1(\tau)$ is:
\begin{eqnarray}
U_{1,(1,1)}^{S_x}\left(\frac{\tau}{4}\right)
&:& \left\{
\begin{array}{lcl}
S_x &\to& \frac{1}{C_1}\left\{\frac{3}{2}C_1C_3\frac{\tau}{4} +[C_2+C_1S_x]^{\frac{3}{2}}
\right\}^{\frac{2}{3}}\\
&-&\frac{C_2}{C_1}\;,
\end{array}
\right. \nonumber\\  \\
U_{2,(1,1)}^{S_x}\left(\frac{\tau}{2}\right)
&:& \left\{
S_x \to S_x - \frac{\tau}{2}  S_y \left(S_z-c_2b\right)
\right.\;,  \nonumber\\\\
U_{1,(1,1)}^{S_x}\left(\frac{\tau}{4}\right)
&:& \left\{
\begin{array}{lcl}
S_x &\to&\frac{1}{C_1}\left\{\frac{3}{2}C_1C_3\frac{\tau}{4} +[C_2+C_1S_x]^{\frac{3}{2}}
\right\}^{\frac{2}{3}}\\
&-&\frac{C_2}{C_1} \;,
\end{array}
\right. \nonumber\\ \\
U_{(1,1)}^{S_y}\left(\frac{\tau}{2}\right)
&:& \left\{
\begin{array}{lcl}
S_y &\to& S_y +\frac{\tau}{2}
\Big[
\frac{\mu S_x(b+\mu S_z)-\mu(\Omega+S_x)S_z}{\sqrt{C^2+2\mu(\Omega S_x - c_1 b S_z)}}
\\
&+&S_x\left(S_z-c_2b\right)
\Big]\;,
\end{array}
\right. \nonumber\\ \\
U_{(1,1)}^{S_z}(\tau)
&:& \left\{
\begin{array}{lcl}
S_z &\to& \frac{B_2}{B_1}
- \frac{1}{B_1}\Big[\left(B_2 - B_1S_z\right)^{\frac{3}{2}}
\\
&-&\frac{3 B_1 B_3\tau}{2}\Big]^{\frac{2}{3}}\;,
\end{array}
\right. \nonumber\\ \\
U_{(1,1)}^{S_y}\left(\frac{\tau}{2}\right)
&:& \left\{
\begin{array}{lcl}
S_y &\to &S_y +\frac{\tau}{2}
\Big[
\frac{\mu S_x(b+\mu S_z)-\mu(\Omega+S_x)S_z}{\sqrt{C^2+2\mu(\Omega S_x - c_1 b S_z)}}
\\
&+&S_x\left(S_z-c_2b\right)
\Big]\;,
\end{array}
\right. \nonumber\\ \\
U_{2,(1,1)}^{S_x}\left(\frac{\tau}{2}\right)
&:& \left\{
S_x \to S_x - \frac{\tau}{2} S_y \left(S_z-c_2b\right)
\right.\;,\\
U_{1,(1,1)}^{S_x}\left(\frac{\tau}{4}\right)
&:& \left\{
\begin{array}{lcl}
S_x &\to &\frac{1}{C_1}\left\{\frac{3}{2}C_1C_3\frac{\tau}{4} +[C_2+C_1S_x]^{\frac{3}{2}}
\right\}^{\frac{2}{3}}\\
&-&\frac{C_2}{C_1}\;,
\end{array}
\right.\nonumber \\ \\
U_{1,(1,1)}^{S_x}\left(\frac{\tau}{4}\right)
&:& \left\{
\begin{array}{lcl}
S_x &\to& \frac{1}{C_1}\left\{\frac{3}{2}C_1C_3\frac{\tau}{4} +[C_2+C_1S_x]^{\frac{3}{2}}
\right\}^{\frac{2}{3}}\\
&-&\frac{C_2}{C_1}\;.
\end{array}
\right. \nonumber\\
\end{eqnarray}

The algorithm provided by $U_{(1,1)}^2(\tau)$ is:
\begin{eqnarray}
U_{(1,1)}^{S_z}\left(\frac{\tau}{2}\right)
&:&\left\{
\begin{array}{lcl}
S_z &\to& \frac{B_2}{B_1}
\\
&-& \frac{1}{B_1}\left\{\left[B_2 - B_1S_z\right]^{\frac{3}{2}}
-\frac{3 B_1 B_3\tau}{4}\right\}^{\frac{2}{3}} \;,
\end{array}
\right.\nonumber\\ \\
U_{1,(1,1)}^{S_x}\left(\frac{\tau}{4}\right)
&:&\left\{
\begin{array}{lcl}
S_x &\to&\frac{1}{C_1}\left\{\frac{3}{2}C_1C_3\frac{\tau}{4} +[C_2+C_1S_x]^{\frac{3}{2}}
\right\}^{\frac{2}{3}}\\
&-&\frac{C_2}{C_1} \;,
\end{array}
\right.\nonumber\\ \\
U_{2,(1,1)}^{S_x}\left(\frac{\tau}{2}\right)
&:&\left\{
S_x \to S_x - \frac{\tau}{2} S_y \left(S_z-c_2b\right)
\right.\;,\\
U_{1,(1,1)}^{S_x}\left(\frac{\tau}{4}\right)
&:&\left\{
\begin{array}{lcl}
S_x &\to&\frac{1}{C_1}\left\{\frac{3}{2}C_1C_3\frac{\tau}{4} +[C_2+C_1S_x]^{\frac{3}{2}}
\right\}^{\frac{2}{3}}\\
&-&\frac{C_2}{C_1} \;,
\end{array}
\right.\nonumber\\ \\
U_{(1,1)}^{S_y}(\tau)
&:&\left\{
\begin{array}{lcl}
S_y &\to& S_y +\tau
\Bigg[
\frac{\mu S_x(b+\mu S_z)-\mu(\Omega+S_x)S_z}{\sqrt{C^2+2\mu(\Omega S_x - c_1 b S_z)}}
\\
&+&S_x\left(S_z-c_2b\right)
\Bigg]\;,
\end{array}
\right.\nonumber\\ \\
U_{1,(1,1)}^{S_x}\left(\frac{\tau}{4}\right)
&:&\left\{
\begin{array}{lcl}
S_x&\to &\frac{1}{C_1}\left\{\frac{3}{2}C_1C_3\frac{\tau}{4} +[C_2+C_1S_x]^{\frac{3}{2}}
\right\}^{\frac{2}{3}}
\\
&-&\frac{C_2}{C_1} \;,
\end{array}
\right.\nonumber\\ \\
U_{2,(1,1)}^{S_x}\left(\frac{\tau}{2}\right)
&:&\left\{
S_x \to S_x - \frac{\tau}{2} S_y \left(S_z-c_2b\right)
\right.\;,\\
U_{1,(1,1)}^{S_x}\left(\frac{\tau}{4}\right)
&:&\left\{
\begin{array}{lcl}
S_x &\to&\frac{1}{C_1}\left\{\frac{3}{2}C_1C_3\frac{\tau}{4} +[C_2+C_1S_x]^{\frac{3}{2}}
\right\}^{\frac{2}{3}}\\
&-&\frac{C_2}{C_1}\;,
\end{array}
\right.\nonumber\\ \\
U_{(1,1)}^{S_z}\left(\frac{\tau}{2}\right)
&:&\left\{
\begin{array}{lcl}
S_z &\to& \frac{B_2}{B_1} \\
&-&\frac{1}{B_1}\left\{\left[B_2 - B_1S_z\right]^{\frac{3}{2}}
-\frac{3 B_1 B_3\tau}{4}\right\}^{\frac{2}{3}} \;.
\end{array}
\right.\nonumber\\
\end{eqnarray}

The algorithm provided by $U_{(1,1)}^3(\tau)$ is:
\begin{eqnarray}
U_{(1,1)}^{S_y}\left(\frac{\tau}{4}\right)
&:&\left\{
\begin{array}{lcl}
S_y &\to&S_y +\frac{\tau}{4}
\Bigg[
\frac{\mu S_x(b+\mu S_z)-\mu(\Omega+S_x)S_z}{\sqrt{C^2+2\mu(\Omega S_x - c_1 b S_z)}}
\\
&+&S_x\left(S_z-c_2b\right) \Bigg] \;,
\end{array}
\right.\nonumber \\\\
U_{(1,1)}^{S_z}\left(\frac{\tau}{2}\right)
&:&\left\{
\begin{array}{lcl}
S_z &\to&\frac{B_2}{B_1} \\
&-&\frac{1}{B_1}\left[\left(B_2 - B_1S_z\right)^{\frac{3}{2}}
-\frac{3 B_1 B_3\tau}{4}\right]^{\frac{2}{3}} \;,
\end{array}
\right.\nonumber \\ \\
U_{(1,1)}^{S_y}\left(\frac{\tau}{4}\right)
&:&\left\{
\begin{array}{lcl}
S_y &\to&S_y +\frac{\tau}{4}
\Bigg[
\frac{\mu S_x(b+\mu S_z)-\mu(\Omega+S_x)S_z}{\sqrt{C^2+2\mu(\Omega S_x - c_1 b S_z)}}
\\
&+&S_x\left(S_z-c_2b\right) \Bigg]\;,
\end{array}
\right.\nonumber\\ \\
U_{1,(1,1)}^{S_x}\left(\frac{\tau}{2}\right)
&:&\left\{
\begin{array}{lcl}
S_x &\to&\frac{1}{C_1}\left\{\frac{3}{2}C_1C_3\frac{\tau}{2} +[C_2+C_1S_x]^{\frac{3}{2}}
\right\}^{\frac{2}{3}}\\
&-&\frac{C_2}{C_1}\;,
\end{array}
\right.\nonumber\\ \\
U_{2,(1,1)}^{S_x}(\tau)
&:&\left\{
S_x \to S_x - \tau S_y\left(S_z-c_2b\right)
\right.\;,\\
U_{1,(1,1)}^{S_x}\left(\frac{\tau}{2}\right)
&:&\left\{
\begin{array}{lcl}
S_x&\to&\frac{1}{C_1}\left\{\frac{3}{2}C_1C_3\frac{\tau}{2} +[C_2+C_1S_x]^{\frac{3}{2}}
\right\}^{\frac{2}{3}}\\
&-&\frac{C_2}{C_1}\;,
\end{array}
\right.\nonumber\\ \\
U_{(1,1)}^{S_y}\left(\frac{\tau}{4}\right)
&:&\left\{
\begin{array}{lcl}
S_y&\to&S_y +\frac{\tau}{4}
\Bigg[ \frac{\mu S_x(b+\mu S_z)
-\mu(\Omega+S_x)S_z}{\sqrt{C^2+2\mu(\Omega S_x - c_1 b S_z)}}
\\
&+&S_x\left(S_z-c_2b\right) \Bigg]\;,
\end{array}
\right.\nonumber\\ \\
U_{(1,1)}^{S_z}\left(\frac{\tau}{2}\right)
&:&\left\{
\begin{array}{lcl}
S_z&\to&\frac{B_2}{B_1} \\
&-&\frac{1}{B_1}\left[\left(B_2 - B_1S_z\right)^{\frac{3}{2}}
-\frac{3 B_1 B_3\tau}{4}\right]^{\frac{2}{3}}\;,
\end{array}
\right.\nonumber\\ \\
U_{(1,1)}^{S_y}\left(\frac{\tau}{4}\right)
&:&\left\{
\begin{array}{lcl}
S_y&\to&S_y +\frac{\tau}{4}
\Bigg[
\frac{\mu S_x(b+\mu S_z)-\mu(\Omega+S_x)S_z}{\sqrt{C^2+2\mu(\Omega S_x - c_1 b S_z)}}
\\
&+&S_x\left(S_z-c_2b\right)
\Bigg]\;.
\end{array}
\right.\nonumber\\ 
\end{eqnarray}

\section{Integration algorithm on the $(2,2)$ surface}
\label{app:int-22}

The algorithm provided by $U_{(2,2)}^1(\tau)$ is
\begin{eqnarray}
U_{1,(2,2)}^{S_x}\left(\frac{\tau}{4}\right)
&:&\left\{
\begin{array}{lcl}
S_x &\to &
\frac{1}{C_1}\left[-\frac{3}{2}C_1C_3\frac{\tau}{4} +\left(C_2+C_1S_x\right)^{\frac{3}{2}}
\right]^{\frac{2}{3}}\\
&-&\frac{C_2}{C_1}\;,
\end{array}
\right.\nonumber\\ \\
U_{2,(2,2)}^{S_x}\left(\frac{\tau}{2}\right)
&:&\left\{
S_x \to S_x -\frac{\tau}{2}  S_y \left(S_z-c_2b\right)
\right.\;,\\
U_{1,(2,2)}^{S_x}\left(\frac{\tau}{4}\right)
&:&\left\{
\begin{array}{lcl}
S_x&\to&
\frac{1}{C_1}\left[-\frac{3}{2}C_1C_3\frac{\tau}{4} +\left(C_2+C_1S_x\right)^{\frac{3}{2}}
\right]^{\frac{2}{3}}\\
&-&\frac{C_2}{C_1}\;,
\end{array}
\right.\nonumber\\ \\
U_{(2,2)}^{S_y}\left(\frac{\tau}{2}\right)
&:&\left\{
\begin{array}{lcl}
S_y&\to&S_y + \frac{\tau}{2}\Bigg[
\frac{\mu(\Omega+S_x)S_z-\mu S_x(b+\mu S_z)}{\sqrt{C^2+2\mu(\Omega S_x - c_1 b S_z)}}
\\
&+&S_x\left(S_z-c_2b\right) \Bigg]\;,
\end{array}
\right.\nonumber\\ \\
U_{(2,2)}^{S_z}(\tau)
&:&\left\{
\begin{array}{lcl}
S_z &\to& \frac{B_2}{B_1} \\
&-&\frac{1}{B_1}\left[\left(B_2-B_1S_z\right)^{\frac{3}{2}}
+ \frac{3}{2} B_1B_3 \tau\right]^{\frac{2}{3}} \;,
\end{array}
\right.\nonumber\\ \\
U_{(2,2)}^{S_y}\left(\frac{\tau}{2}\right)
&:&\left\{
\begin{array}{lcl}
S_y &\to &S_y + \frac{\tau}{2}\Bigg[
\frac{\mu(\Omega+S_x)S_z-\mu S_x(b+\mu S_z)}{\sqrt{C^2+2\mu(\Omega S_x - c_1 b S_z)}}
\\
&+&S_x\left(S_z-c_2b\right) \Bigg]\;,
\end{array}
\right.\nonumber\\ \\
U_{1,(2,2)}^{S_x}\left(\frac{\tau}{4}\right)
&:&\left\{
\begin{array}{lcl}
S_x &\to&
\frac{1}{C_1}\left[-\frac{3}{2}C_1C_3\frac{\tau}{4} +\left(C_2+C_1S_x\right)^{\frac{3}{2}}
\right]^{\frac{2}{3}}\\
&-&\frac{C_2}{C_1}\;,
\end{array}
\right.\nonumber\\ \\
U_{2,(2,2)}^{S_x}\left(\frac{\tau}{2}\right)
&:&\left\{
S_x \to S_x -\frac{\tau}{2} S_y\left(S_z-c_2b\right)
\right.\;,\\
U_{1,(2,2)}^{S_x}\left(\frac{\tau}{4}\right)
&:&\left\{
\begin{array}{lcl}
S_x  &\to &
\frac{1}{C_1}\left[-\frac{3}{2}C_1C_3\frac{\tau}{4} +\left(C_2+C_1S_x\right)^{\frac{3}{2}}
\right]^{\frac{2}{3}}\\
&-&\frac{C_2}{C_1}\;.
\end{array}
\right.\nonumber\\ 
\end{eqnarray}

The algorithm provided by $U_{(2,2)}^2(\tau)$ is
\begin{eqnarray}
U_{(2,2)}^{S_z}\left(\frac{\tau}{2}\right)
&:&\left\{
\begin{array}{lcl}
S_z &\to& \frac{B_2}{B_1} \\
&-&\frac{1}{B_1}\left[\left(B_2-B_1S_z\right)^{\frac{3}{2}}
+ \frac{3}{2} B_1B_3 \frac{\tau}{2}\right]^{\frac{2}{3}} \;,
\end{array}
\right.\nonumber\\ \\
U_{1,(2,2)}^{S_x}\left(\frac{\tau}{4}\right)
&:&\left\{
\begin{array}{lcl}
S_x &\to&
\frac{1}{C_1}\left[-\frac{3}{2}C_1C_3\frac{\tau}{4} +\left(C_2+C_1S_x\right)^{\frac{3}{2}}
\right]^{\frac{2}{3}}\\
&-&\frac{C_2}{C_1}\;,
\end{array}
\right.\;,\nonumber\\ \\
U_{2,(2,2)}^{S_x}\left(\frac{\tau}{2}\right)
&:&\left\{
S_x \to S_x -\frac{\tau}{2} S_y \left(S_z-c_2b\right)
\right.\;,\\
U_{1,(2,2)}^{S_x}\left(\frac{\tau}{4}\right)
&:&\left\{
\begin{array}{lcl}
S_x&\to&
\frac{1}{C_1}\left[-\frac{3}{2}C_1C_3\frac{\tau}{4} +\left(C_2+C_1S_x\right)^{\frac{3}{2}}
\right]^{\frac{2}{3}}\\
&-&\frac{C_2}{C_1} \;,
\end{array}
\right.\nonumber\\ \\
U_{(2,2)}^{S_y}(\tau)
&:&\left\{
\begin{array}{lcl}
S_y &\to& S_y + \tau\Bigg[
\frac{\mu(\Omega+S_x)S_z-\mu S_x(b+\mu S_z)}{\sqrt{C^2+2\mu(\Omega S_x - c_1 b S_z)}}
\\
&+&S_x\left(S_z-c_2b\right) \Bigg]\;,
\end{array}
\right.\nonumber\\ \\
U_{1,(2,2)}^{S_x}\left(\frac{\tau}{4}\right)
&:&\left\{
\begin{array}{lcl}
S_x &\to&
\frac{1}{C_1}\left[-\frac{3}{2}C_1C_3\frac{\tau}{4} +\left(C_2+C_1S_x\right)^{\frac{3}{2}}
\right]^{\frac{2}{3}}\\
&-&\frac{C_2}{C_1} \;,
\end{array}
\right.\nonumber\\ \\
U_{2,(2,2)}^{S_x}\left(\frac{\tau}{2}\right)
&:&\left\{
S_x \to S_x -\frac{\tau}{2} S_y\left(S_z-c_2b\right)
\right.\;,\\
U_{1,(2,2)}^{S_x}\left(\frac{\tau}{4}\right)
&:&\left\{
\begin{array}{lcl}
S_x &\to&
\frac{1}{C_1}\left[-\frac{3}{2}C_1C_3\frac{\tau}{4} +\left(C_2+C_1S_x\right)^{\frac{3}{2}}
\right]^{\frac{2}{3}}\\
&-&\frac{C_2}{C_1}\;,
\end{array}
\right.\nonumber\\ \\
U_{(2,2)}^{S_z}\left(\frac{\tau}{2}\right)
&:&\left\{
\begin{array}{lcl}
S_z &\to& \frac{B_2}{B_1} \\
&-&\frac{1}{B_1}\left[\left(B_2-B_1S_z\right)^{3/2}
+ \frac{3}{2} B_1B_3 \frac{\tau}{2}\right]^{2/3}\;.
\end{array}
\right.\nonumber\\ 
\end{eqnarray}

The algorithm provided by $U_{(2,2)}^3(\tau)$ is
\begin{eqnarray}
U_{(2,2)}^{S_y}\left(\frac{\tau}{4}\right)
&:&\left\{
\begin{array}{lcl}
S_y &\to &S_y + \frac{\tau}{4}\Bigg[
\frac{\mu(\Omega+S_x)S_z-\mu S_x(b+\mu S_z)}{\sqrt{C^2+2\mu(\Omega S_x - c_1 b S_z)}}
\\
&+&S_x\left(S_z-c_2b\right) \Bigg] \;,
\end{array}
\right.\nonumber\\ \\
U_{(2,2)}^{S_z}\left(\frac{\tau}{2}\right)
&:&\left\{
\begin{array}{lcl}
S_z& \to& \frac{B_2}{B_1} \\
&-&\frac{1}{B_1}\left[\left(B_2-B_1S_z\right)^{\frac{3}{2}}
+ \frac{3}{2} B_1B_3 \frac{\tau}{2}\right]^{\frac{2}{3}} \;,
\end{array}
\right.\nonumber\\ \\
U_{(2,2)}^{S_y}\left(\frac{\tau}{4}\right)
&:&\left\{
\begin{array}{lcl}
S_y &\to& S_y + \frac{\tau}{4}\Bigg[
\frac{\mu(\Omega+S_x)S_z-\mu S_x(b+\mu S_z)}{\sqrt{C^2+2\mu(\Omega S_x - c_1 b S_z)}}
\\
&+&S_x \left(S_z-c_2b\right)\Bigg]\;,
\end{array}
\right.\nonumber\\ \\
U_{1,(2,2)}^{S_x}\left(\frac{\tau}{2}\right)
&:&\left\{
\begin{array}{lcl}
S_x &\to&
\frac{1}{C_1}\left[-\frac{3}{2}C_1C_3\frac{\tau}{2} +\left(C_2+C_1S_x\right)^{\frac{3}{2}}
\right]^{\frac{2}{3}}\\
&-&\frac{C_2}{C_1} \;,
\end{array}
\right.\nonumber\\ \\
U_{2,(2,2)}^{S_x}(\tau)
&:&\left\{
S_x \to S_x -\tau S_y\left(S_z-c_2b\right)
\right.\;,\\
U_{1,(2,2)}^{S_x}\left(\frac{\tau}{2}\right)
&:&\left\{
\begin{array}{lcl}
S_x & \to&
\frac{1}{C_1}\left[-\frac{3}{2}C_1C_3\frac{\tau}{2} +\left(C_2+C_1S_x\right)^{\frac{3}{2}}
\right]^{\frac{2}{3}}\\
&-&\frac{C_2}{C_1}\;,
\end{array}
\right.\nonumber\\ \\
U_{(2,2)}^{S_y}\left(\frac{\tau}{4}\right)
&:&\left\{
\begin{array}{lcl}
S_y &\to& S_y + \frac{\tau}{4}\Bigg[
\frac{\mu(\Omega+S_x)S_z-\mu S_x(b+\mu S_z)}{\sqrt{C^2+2\mu(\Omega S_x - c_1 b S_z)}}
\\
&+&S_x\left(S_z-c_2b\right) \Bigg]\;,
\end{array}
\right.\nonumber\\ \\
U_{(2,2)}^{S_z}\left(\frac{\tau}{2}\right)
&:&\left\{
\begin{array}{lcl}
S_z &\to& \frac{B_2}{B_1}\\
&-&\frac{1}{B_1}\left[\left(B_2-B_1S_z\right)^{\frac{3}{2}}
+ \frac{3}{2} B_1B_3 \frac{\tau}{2}\right]^{\frac{2}{3}} \;,
\end{array}
\right.\nonumber\\ \\
U_{(2,2)}^{S_y}\left(\frac{\tau}{4}\right)
&:&\left\{
\begin{array}{lcl}
S_y &\to& S_y + \frac{\tau}{4}\Bigg[
\frac{\mu(\Omega+S_x)S_z-\mu S_x(b+\mu S_z)}{\sqrt{C^2+2\mu(\Omega S_x - c_1 b S_z)}}
\\
&+&S_x\left(S_z-c_2b\right) \Bigg]\;.
\end{array}
\right.\nonumber\\ 
\end{eqnarray}

\section{Integration algorithm on the $(1,2)$ surface}
\label{app:int-12}

The algorithm provided by $U_{(1,2)}^1(\tau)$ is
\begin{eqnarray}
U_{(1,2)}^{S_x}\left(\frac{\tau}{2}\right) &:&
\left\{ S_x \to S_x -  \frac{\tau}{2}S_y\left(S_z-c_2b\right) 
\right.\;,
\\
U_{(1,2)}^{S_y}(\tau) &:& \left\{  S_y \to S_y +\tau S_x\left(S_z-c_2b\right) 
\right.\;,
\\
U_{(1,2)}^{S_x}\left(\frac{\tau}{2}\right) &:&
\left\{ S_x \to S_x - \frac{\tau}{2}S_y\left(S_z-c_2b\right)
\right.\;.
\end{eqnarray}

The algorithm provided by $U_{(1,2)}^2(\tau)$ is
\begin{eqnarray}
U_{(1,2)}^{S_y}\left(\frac{\tau}{2}\right) &:&
\left\{  S_y \to S_y +\frac{\tau}{2}  S_x\left(S_z-c_2b\right)
\right.\;,
\\
U_{(1,2)}^{S_x}(\tau) &:& \left\{ S_x \to S_x - \tau S_y\left(S_z-c_2b\right)
\right.\;,
\\
U_{(1,2)}^{S_y}\left(\frac{\tau}{2}\right) &:&
 \left\{  S_y \to S_y +\frac{\tau}{2}  S_x\left(S_z-c_2b\right)
\right.\;.
\end{eqnarray}


\section*{Acknowledgements}
The author is grateful to G. S. Ezra for the many discussions, encouragement
and support received in the past few years.

This work is based upon research supported by
the National Research Foundation of South Africa.



\begin{thebibliography}{}
\bibitem{kapral}
R. Kapral and A. Sergi,
Dynamics of Condensed Phase Proton and Electron Transfer Processes,
in {\em Handbook of Theoretical and Computational Nanotechnology},
Vol. 1, Ch. 92. Eds. M. Rieth and W. Schommers.
American Scientific Publishers, Valencia CA (2005).
\bibitem{josephson}
K. K. Likharev, Dynamics of Josephson Junctions and Circuits.
CRC Press, Amsterdam (1996).
\bibitem{dots1}
T. Hayashi, T. Fujisawa, H. D. Cheong, Y. H. Jeong, and Y. Hirayama,
Coherent Manipulation of Electronic States in a Double Quantum Dot,
Phys. Rev. Lett., {\bf 91}, 226804 4pp (2003).
\bibitem{dots2}
J. R. Petta, A. C. Johnson, C. M. Marcus, M. P. Hanson, and A. C. Gossard,
Manipulation of a single charge in a double quantum dot,
Phys. Rev. Lett., {\bf 93}, 186802 4pp (2004).
\bibitem{dots3}
J. R. Petta, A. C. Johnson, J. M. Taylor, E. A. Laird, A. Yacoby, M. D. Lukin, C. M. Marcus,
M. P. Hanson, and A. C. Gossard,
Coherent manipulation of coupled electron spins in semiconductor quantum dots,
Science {\bf 309}, 2180-2184 (2005).
\bibitem{leggett}
A. J. Leggett, S. Chakravarty, A. T. Dorsey, M. P. A. Fisher, A. Garg, and
W. Zwerger, 
Dynamics of the dissipative two-state system,
Rev. Mod. Phys., {\bf 59}, 1-85 (1987).
\bibitem{prokofev}
N. V. Prokof'ev and P. C. E. Stamp,
Theory of the spin bath,
Rep. Prog. Phys., {\bf 63}, 669-726 (2000).
\bibitem{petruccione}
H.-P. Breuer and F.Petruccione, The Theory
of Open Quantum Systems. Oxford University Press, Oxford (2002).
\bibitem{ottinger}
H. C. Ottinger, Beyond Equilibrium Thermodynamics.
John Wiley \& Sons, Hoboken (2005).
\bibitem{ottinger-qm}
H. C. Ottinger, 
Derivation of a two-generator framework of nonequilibrium
thermodynamics for quantum system,
Phys. Rev. E, {\bf 62}, 4720 5pp (2000).
\bibitem{ottinger-qm2}
H. C. Ottinger,
Nonlinear thermodynamic quantum master equation: Properties and examples,
Phys.Rev. A, {\bf 82}, 052119 11pp (2010),
\bibitem{ottinger-qm3}
H. C. Ottinger,
Euro Physics Letters, {\bf 94}, 10006 6pp (2011).
\bibitem{ottinger-qm4}
H. C. Ottinger, 
Stochastic process behind nonlinear thermodynamic quantum master equation. 
I. Mean-field construction,
Phys. Rev. A, {\bf 86}, 032101 5pp (2012).
\bibitem{ottinger-qm5}
J. Flakowski, M. Schweizer, and H. C. Ottinger,
Stochastic process behind nonlinear thermodynamic quantum master equation. II. Simulation,
Phys. Rev. A, {\bf 86}, 032102 9pp (2012).
\bibitem{sergi-spin}
A. Sergi, 
Communication: Quantum dynamics in classical spin baths,
J. Chem. Phys., {\bf 139}, 031101 4pp (2013).
\bibitem{anderson}
A. Anderson, Quantum Backreaction on ``Classical'' Variables,
Phys. Rev. Lett. {\bf 74} 621-625 (1995).
\bibitem{prezhdo}
O. Prezhdo and V. V. Kisil, 
Mixing Quantum and Classical Mechanics,
Phys. Rev. A, {\bf 56}, 162-176 (1997).
\bibitem{balescu}
W. Y. Zhang and R. Balescu, 
Statistical mechanics of a spin polarized plasma,
J. Plasma Phys., {\bf 40}, 199-213 (1988).
\bibitem{balescu2}
R. Balescu and W. Y. Zhang, 
Kinetic equation, spin hydrodynamics and collisional depolarization
rate in a spin-polarized plasma,
J. Plasma Physics, {\bf 40}, 215-234 (1988).
\bibitem{mcquarrie}
T. A. Osborn, M. F. Kondrat'eva, G. C. Tabisz, and B. R. McQuarrie,
Mixed Weyl symbol calculus and spectral line shape theory,
J. Phys. A, {\bf 32}, 4149-4169 (1999).
\bibitem{qcl1}
V. I. Gerasimenko,
Dynamical equations of quantum-classical systems,
Teoret. Mat. Fiz., {\bf 50},  77–87 (1982).
\bibitem{qcl2}
W. Boucher and J. Traschen, 
Semiclassical physics and quantum fluctuations,
Phys. Rev. D, {\bf 37}, 3522-3532 (1988).
\bibitem{qcl3}
C. C. Martens and and J.-Y. Fang,
Semiclassical limit molecular dynamics on multiple electronic surfaces,
J. Chem. Phys., {\bf 106}, 4918-4930 (1996).
\bibitem{qcl4}
R. Kapral and G. Ciccotti, Mixed quantum-classical dynamics,
J. Chem. Phys., {\bf 110}, 8919-8929 (1999).
\bibitem{qcl5}
I. Horenko, C. Salzmann, B. Schmidt, and C. Schutte,
Quantum-classical Liouville approach to molecular dynamics:
Surface hopping Gaussian phase-space packets,
J. Chem. Phys., {\bf 117}, 11075-11088 (2002).
\bibitem{qcl6}
Q. Shi and E. Geva,
A derivation of the mixed quantum-classical
Liouville equation from the influence functional formalism,
J. Chem. Phys.,  {\bf 121}, 3393-3404 (2004).
\bibitem{ilya}
A. Sergi, I. Sinayskiy, and F. Petruccione,
Numerical and Analytical Approach to the Quantum Dynamics
of Two Coupled Spins in Bosonic Baths,
Physical Review A, {\bf 80}, 012108 7pp (2009).
\bibitem{allentildesley}
M. P. Allen and D. J. Tildesley,
Computer Simulation of Liquids. Oxford University Press, Oxford (2009).
\bibitem{frenkelsmit}
D. Frenkel and B. Smit,
Understanding Molecular Simulation. Academic Press, London (2002).
\bibitem{b3}
A. Sergi, 
Non-Hamiltonian Commutators in Quantum Mechanics,
Phys. Rev. E, {\bf 72}, 066125 9pp (2005).
\bibitem{b4}
A. Sergi, 
Deterministic constant-temperature dynamics for dissipative quantum systems,
J. Phys. A: Math. Theor., {\bf 40}, F347-F354 (2007).
\bibitem{b-silurante}
A. Sergi,
Statistical Mechanics of Quantum-Classical Systems with Holonomic Constraints,
J. Chem. Phys., {\bf 124}, 024110 10pp (2006).
\bibitem{b1}
A. Sergi and M. Ferrario,
Non-Hamiltonian  Equations of Motion with a Conserved Energy,
Phys. Rev. E, {\bf 64}, 056125 9pp (2001).
\bibitem{b2}
A. Sergi,
Non-Hamiltonian Equilibrium Statistical Mechanics,
Phys. Rev. E, {\bf 67}, 021101 7pp (2003).
\bibitem{sergi-pvg}
A. Sergi and P. V. Giaquinta,
On the geometry and entropy of non-Hamiltonian phase space,
Journal of Statistical Mechanics: Theory and Experiment, {\bf 02}, P02013 20pp (2007).
\bibitem{holo1}
A. Sergi,
Generalized Bracket Formulation of Constrained Dynamics in Phase Space,
Phys. Rev. E, {\bf 69}, 021109 9pp (2004).
\bibitem{holo2}
A. Sergi,
Phase Space Flows for Non-Hamiltonian Systems with Constraints,
Phys. Rev. E, {\bf 72}, 031104 5pp (2005).
\bibitem{gonzalo}
G. G. de Polavieja and E. Sj\"oqvist, 
Extending the quantal adiabatic theorem: Geometry of noncyclic motion,
Am. J. Phys., {\bf 66}, 431-438 (1998).
\bibitem{truhlar}
C. A. Mead and D. G. Truhlar, 
On the determination of Born-Oppenheimer nuclear motion
wave functions including complications due to conical intersections
and identical nuclei,
J. Chem. Phys., {\bf 70}, 2284-2296 (1979).
\bibitem{berry}
M. V. Berry, 
Quantal Phase Factors Accompanying Adiabatic Cjanges,
Proc. R. Soc. London, Ser. A, {\bf 392}, 45-57 (1984).
\bibitem{qphases}
Geometric Phases in Physics, eds. A. Shapere and F. Wilczek.
World Scientific, Singapore (1989).
\bibitem{mead}
C. A. Mead, 
The geometric phase in molecular systems,
Rev. Mod. Phys., {\bf 64}, 51-85 (1992).
\bibitem{kuratsuji}
H. Kuratsuji and S. Iida, 
Effective Action for Adiabatic Process,
Prog. Theor. Phys., {\bf 74}, 439-445 (1985).
\bibitem{respa}
M. Tuckerman, G. J. Martyna, and B. J. Berne, 
Reversible multiple time scale molecular dynamics,
J. Chem. Phys., {\bf 97}, 1990-2001 (1992).
\bibitem{respa2}
G. J. Martyna, M. Tuckerman, D. J. Tobias, and
M. L. Klein, 
Explicit Reversible integrators for extended systems dynamics,
Molec. Phys., {\bf 87}, 1117-1157 (1996).
\bibitem{ribes}
A. Sergi, M. Ferrario, and D. Costa, 
Reversible Integrators for Basic Extended System Molecular Dynamics,
Mol. Phys., {\bf 97}, 825-832 (1999).
\bibitem{ezra}
G. S. Ezra, 
Reversible measure-preserving integrators for non-Hamiltonian systems,
J. Chem. Phys., {\bf 125}, 034104 14pp (2006). 
\bibitem{sergi-ezra}
A. Sergi and G. S. Ezra, 
Bulgac-Kusnezov-Nose-Hoover thermostats,
Phys. Rev. E, {\bf 81}, 036705 14pp (2010).
\bibitem{sergi-ezra2}
A. Sergi and G. S. Ezra, Bulgac-Kusnezov-Nos\'e-Hoover thermostats for spins,
\emph{unpublished}.
\bibitem{nielsen}
S. Nielsen, R. Kapral, and G. Ciccotti, 
Statistical mechanics of quantum-classical systems,
J. Chem. Phys.,  {\bf 115}, 5805-5815 (2001).
\bibitem{pe}
A. Sergi and P. V. Giaquinta, 
On computational strategies in molecular dynamics simulation,
Physics Essays, {\bf 20}, 629-640 (2007).
\bibitem{pati}
A. K. Pati,
Adiabatic berry phase and hannay angle for open paths,
Ann. Phys., {\bf 270} 178-197 (1998).
\bibitem{filipp}
S. Filipp and E. Sj\"oqvist,
Off-diagonal generalization of the mixed-state geometric phase
Phys. Rev. A, {\bf 68}, 042112, 10pp (2003).
\bibitem{englman}
R. Englman, A. Yahalom, and M. Baer,
The open path phase for degenerate and non-degenerate systems and
its relation to the wave function and its modulus,
Eur. Phys. J. D, {\bf 8}, 1-7 (2000).
\bibitem{manini}
N. Manini and F. Pistolesi, Off-diagonal Geometric phases,
Phys. Rev. Lett., {\bf 85}, 3067-3071 (2000).
\bibitem{leimkuhler}
J. Frank, W. Huang, and B. Leimkuhler 	
Geometric integrators for classical spin systems,
J. Comp. Phys., {\bf 133}, 160-172 (1997).
\bibitem{landau}
M. Krech, A. Bunker, and D. P. Landau,
Fast spin dynamics algorithms for classical spin systems,
Comput. Phys. Comm., {\bf 111}, 1-13 (1998).
\bibitem{steinigeweg}
R. Steinigeweg and H.-J. Schmidt,
Symplectic integrators for classical spin systems
Comput. Phys. Comm., {\bf 174}, 853-861 (2006).
\bibitem{yoshida}
H. Yoshida, 
Construction of higher order symplectic integrators,
Phys. Lett. A, {\bf 150}, 262-268 (1990).


\end{thebibliography}
\end{document}